\documentclass[journal]{IEEEtran}
\usepackage{amsmath,amsfonts}
\usepackage{algorithmic}
\usepackage{array}
\usepackage{textcomp}
\usepackage{stfloats}
\usepackage{url}
\usepackage{verbatim}
\usepackage{graphicx}

\def\BibTeX{{\rm B\kern-.05em{\sc i\kern-.025em b}\kern-.08em
    T\kern-.1667em\lower.7ex\hbox{E}\kern-.125emX}}
\usepackage{balance}

\usepackage{booktabs,makecell, multirow, tabularx}
\usepackage{microtype}		
\usepackage{cleveref} 		
\usepackage{setspace} 	    
\usepackage{subfig}

\crefname{figure}{Fig.}{Figs.}
\crefname{table}{Table}{Tables}
\crefname{equation}{Eq.}{Eqs.}

\newcolumntype{L}[1]{>{\raggedright\let\newline\\\arraybackslash\hspace{0pt}}m{#1}}
\newcolumntype{C}[1]{>{\centering\let\newline\\\arraybackslash\hspace{0pt}}m{#1}}
\newcolumntype{R}[1]{>{\raggedleft\let\newline\\\arraybackslash\hspace{0pt}}m{#1}}

\begin{document}

\title{The Application of Deep Learning for Lymph Node Segmentation: A Systematic Review}

\author{Jingguo Qu, Xinyang Han, Man-Lik Chui, Yao Pu, Simon Takadiyi Gunda, Ziman Chen, Jing Qin \IEEEmembership{Senior Member, IEEE}, Ann Dorothy King, Winnie Chiu-Wing Chu, Jing Cai, and Michael Tin-Cheung Ying \thanks{This work was supported by a General Research Fund of the Research Grant Council of Hong Kong (Reference no. 15102222). \textit{(Corresponding authors: Michael Tin-Cheung Ying.)}} \thanks{Jingguo Qu, Xinyang Han, Man-Lik Chui, Yao Pu, Simon Takadiyi Gunda, Ziman Chen, Jing Cai, and Michael Tin-Cheung Ying are with the Department of Health Technology and Informatics, The Hong Kong Polytechnic University, Hong Kong, China (e-mail: jingguo.qu@connect.polyu.hk). } \thanks{Jing Qin is with the Centre for Smart Health and School of Nursing, The Hong Kong Polytechnic University, Hong Kong, China.} \thanks{Ann Dorothy King and Winnie Chiu-Wing Chu are with the Department of Imaging and Interventional Radiology, The Chinese University of Hong Kong, Hong Kong, China.}}

\markboth{preprint}{The Application of Deep Learning for Lymph Node Segmentation: A Systematic Review}

\maketitle

\begin{abstract}
	Automatic lymph node segmentation is the cornerstone for advances in computer vision tasks for early detection and staging of cancer. Traditional segmentation methods are constrained by manual delineation and variability in operator proficiency, limiting their ability to achieve high accuracy. The introduction of deep learning technologies offers new possibilities for improving the accuracy of lymph node image analysis. This study evaluates the application of deep learning in lymph node segmentation and discusses the methodologies of various deep learning architectures such as convolutional neural networks, encoder-decoder networks, and transformers in analyzing medical imaging data across different modalities. Despite the advancements, it still confronts challenges like the shape diversity of lymph nodes, the scarcity of accurately labeled datasets, and the inadequate development of methods that are robust and generalizable across different imaging modalities. To the best of our knowledge, this is the first study that provides a comprehensive overview of the application of deep learning techniques in lymph node segmentation task. Furthermore, this study also explores potential future research directions, including multimodal fusion techniques, transfer learning, and the use of large-scale pre-trained models to overcome current limitations while enhancing cancer diagnosis and treatment planning strategies.
\end{abstract}

\begin{IEEEkeywords}
	Convolutional neural network, deep learning, lymph node segmentation, medical image processing, transformer.
\end{IEEEkeywords}

\section{Introduction}

The lymphatic system is a crucial part of the immune system. It consists of lymph nodes (LNs) which are found in various parts of the body such as the neck, axillae, chest, abdomen, and pelvis. These nodes may change in size and appearance in response to infection and inflammation, as well to metastatic spread from a primary cancer.

Medical imaging technology is pivotal in clinical diagnostics, with the advantage of non-invasive illustration of the body parts. Computed tomography (CT), positron emission tomography (PET), magnetic resonance imaging (MRI), and ultrasonography (US) are common medical imaging methods in LN assessment. The determination of abnormal LNs has always posed a significant challenge in clinical assessment. While most cases of lymphadenopathy are benign, such as reactive hyperplasia, a small proportion are malignant~\cite{lymphadenopathy_bazemore_2002}. These malignant LNs may indicate lymphoma or metastases from other primary malignancies. The diagnosis of a malignant node in these modalities relies heavily on size, shape, necrosis, and extranodal extension. Unfortunately, some of the malignant features may overlap with both normal nodes and those involved in inflammation and infection~\cite{ultrasound_ahuja_2008,sonographic_ahuja_2005}.

In cancer patients, the identification of metastatic LNs requires meticulousness as an inappropriate diagnosis can be detrimental to the patients. However, the current identification process is not only time-consuming and labor-intensive but also subject to variability in accuracy and consistency, depending on the experience and expertise of the physician. Given the aforementioned background, it is therefore imperative to find alternative ways to automatically detect metastatic nodes efficiently, and accurately. Advancements in automatic detection require accurate automatic segmentation of LNs. The automatic segmentation method extracts the region of interest in medical images without manual intervention and could help standardize the process, ensuring uniformity in the evaluation, and significantly reducing the time and effort required by clinicians.

In recent years, deep learning techniques are widely used in image processing such as convolutional neural network (CNN)~\cite{cnn_lecun_1989}, deep residual network (ResNet)~\cite{resnet_kaiming_2016}, U-Net~\cite{unet_ronneberger_2015} and Vision Transformer (ViT)~\cite{vit_dosovitskiy_2021}. These architectures of deep learning are capable of learning the representation of images and automating the identification process. The significant development of deep learning techniques has laid a solid foundation for medical image processing tasks, such as liver lesion classification~\cite{lesion_cls_adar_2018}, liver and vessel segmentation~\cite{vessel_seg_jin_2019,practical_ansari_2022,neural_ansari_2023,towards_ansari_2023}, lung lesion detection~\cite{retinaunet_lin_2017}, low-dose CT image reconstruction~\cite{competitive_shan_2019}, and elastogram generation~\cite{unveiling_ansari_2023}.

Prior studies~\cite{advancements_ansari_2024,deep_rayed_2024,review_aljabri_2022} have provided an overview of the application of deep learning techniques to medical image segmentation tasks across various anatomical sites. Despite the exploration of deep learning techniques for medical image segmentation in various modalities, tissues, and organs in the aforementioned studies, a review addressing LN segmentation tasks remains to be reported. By leveraging advancements in deep learning and medical imaging, automated LN segmentation holds the potential to enhance diagnostic accuracy, streamline clinical workflows, and improve patient outcomes. Deep learning methods have been widely adopted for LN segmentation across different imaging modalities. To the best of our knowledge, this is the first study that provides a comprehensive overview of the application of deep learning techniques in LN segmentation task and highlights their significance in improving the accuracy of LN image analysis. The main contributions of this study are as follows:
\begin{itemize}
	\item We conduct a systematic review of the application of deep learning techniques for LN segmentation among commonly utilized medical imaging modalities.
	\item We analysis and compare the segmentation performance reported in included studies in perspectives of imaging modalities and method architectures.
	\item We discuss the current challenges and limitations appeared in included studies of LN segmentation and provide potential directions for the future research.
\end{itemize}

This study is organized as follows: Section~\ref{sec:method} introduces the method to conduct this systematic review and evaluation metrics for segmentation performance assessment. Section~\ref{sec:result} provides a detailed categorization of different deep learning approaches for LN segmentation. Section~\ref{sec:discussion} concludes with a summary and discussion of the current state of research, identifies the key issues faced, and outlines potential directions for future investigation. Section~\ref{sec:conclusion} summarizes the study.

\section{Method}\label{sec:method}

\begin{figure*}[htb]
	\centering
	\includegraphics[width=0.8\linewidth]{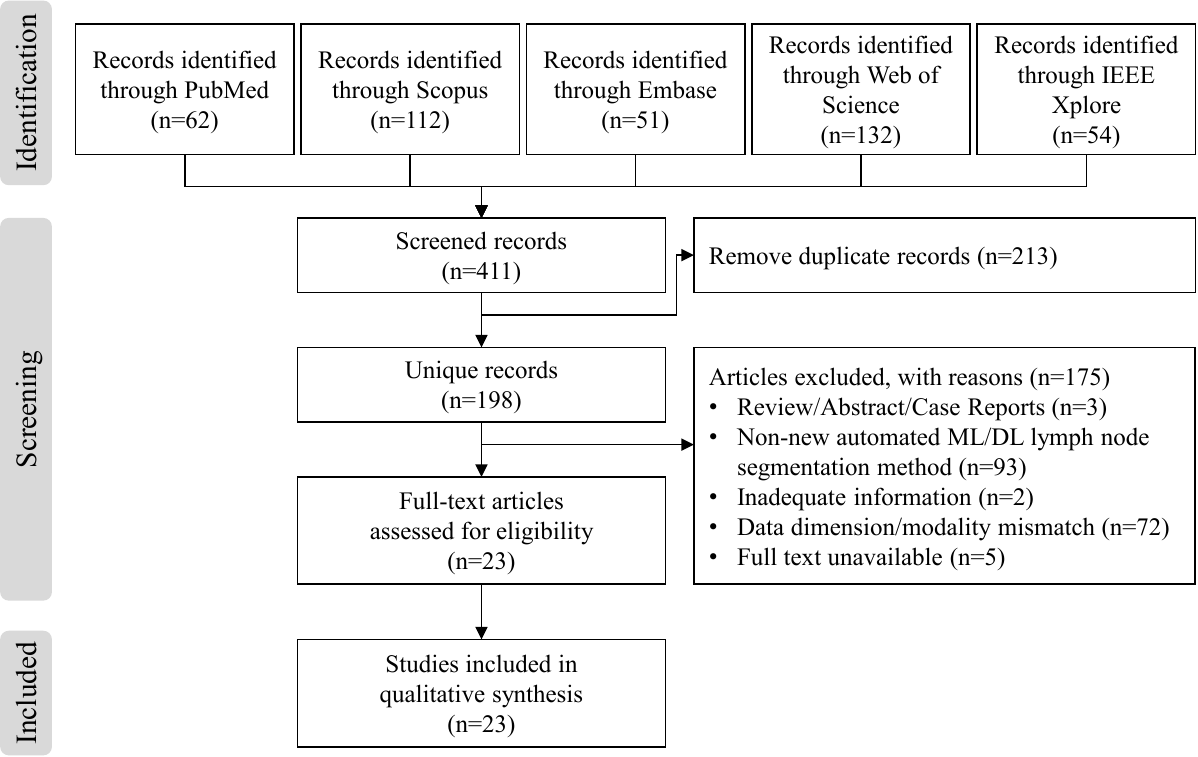}
	\caption{PRISMA systematic review flowchart.}
	\label{fig:screening}
\end{figure*}

This review follows the Preferred Reporting Items for Systematic Reviews and Meta-Analyses (PRISMA) guidelines~\cite{prisma_2020}. Figure~\ref{fig:screening} illustrates the literature search process.

\subsection{Data sources and search strategy}

A systematic search was conducted in PubMed, Scopus, Embase, Web of Science and IEEE Xplore to identify relevant studies published from Jan 2014 to Dec 2024. The same search string was used for all databases, with different syntaxes to match the search requirements of each database. The search terms used in titles and abstracts are: ("segmentation" OR "segment") AND ("artificial intelligence" OR "machine learning" OR "deep learning" OR "neural networks" OR "reinforcement learning" OR "supervised learning" OR "unsupervised learning" OR "CNN" OR "convolutional" OR "transformer" OR "attention" OR "autoencoder") AND ("intersection over union" OR "iou" OR "f1" OR "f1 score" OR "HD" OR "Hausdorff distance" OR "dice" OR "dice score") and ("lymph node").

This search ranged from Jan 2014 to Dec 2024 to ensure the inclusion of the most recent studies and was limited to English-language articles. The search was conducted on 7th Dec 2024.

\subsection{Study selection}

Two reviewers (Qu and Han) independently screened the titles and abstracts of the articles to determine their eligibility for full-text review. The selected articles were then reviewed to determine their eligibility for inclusion in the review. Disagreements between the two reviewers were resolved by discussion with a third reviewer. The Cohen's kappa coefficient was calculated to assess the inter-rater agreement between the two reviewers ($\kappa$=0.793). The detailed selection results and calculation of Cohen's kappa coefficient can be found in supplementary materials.

Studies met the following criteria were included:
\begin{enumerate}
	\item The objective of the study was to segment the single lymph node or the lymph node cluster.
	\item The modality of the image data used for the study belongs to one of the following modalities: computed tomography (CT), positron emission tomography (PET), magnetic resonance imaging (MRI), and ultrasound (US).
	\item Articles that described information related to the dataset, such as the source of the data, the number of data included, the data pre-processing and post-processing methods, and the proportion of data used for model training and testing, etc.
	\item Articles that have clearly stated segmentation results and quantitative assessment indicators are stated such as dice similarity coefficient (DSC), Hausdorff distance (HD), intersection over union (IoU), etc.
\end{enumerate}

The exclusion criteria for the article selection were:
\begin{enumerate}
	\item All review articles, letters, abstracts, and case reports.
	\item The objective of the study was not lymph node segmentation, the methodology used in the study was not related to machine learning/deep learning, no new segmentation methods were proposed, or the segmentation method proposed is not completely automated.
	\item Inadequate information, the detailed information of machine learning/deep learning model is missing, such as model structure, hyper-parameters, loss function, etc.
	\item The images used for segmentation in the article are not in 2D form, or the modality of images mismatches.
	\item Non-English papers or full text unavailable.
\end{enumerate}

Based on the above search and selection strategy, 411 publications were identified. After removing duplicates, 198 publications were included. Among the 198 full-text publications, 175 were removed according to the above exclusion and inclusion criteria. Finally, 23 full-text publications were included in this study.

\subsection{Data extraction}

Key information relevant to the segmentation method was extracted from the included studies. The extracted information included the following:
\begin{enumerate}
	\item The overview of the study and the backbone architecture of the proposed model.
	\item The size, site, modality and source of the image dataset used in the study.
	\item The data augmentation techniques.
	\item The performance evaluation results.
\end{enumerate}

\subsection{Evaluation metrics}

\begin{figure}[htb]
	\centering
	\includegraphics[width=0.3\linewidth]{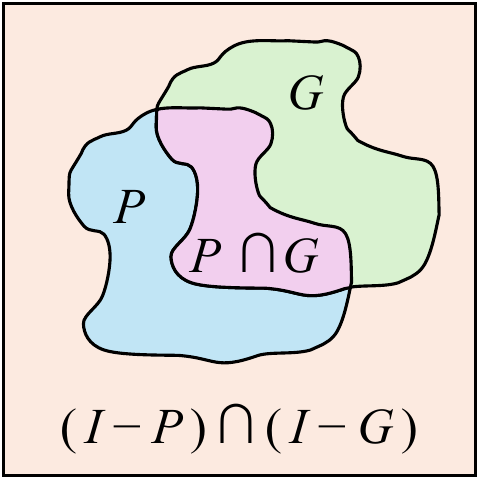}
	\caption{Definition of regions.}
	\label{fig:regions}
\end{figure}

Image segmentation is the process of classifying all pixels in an image into multiple classes, and it is known as binary segmentation when the number of classes equals two (typically the foreground and background). The LN segmentation task is a standard binary segmentation task, where the foreground represents the LNs, and the background indicates other regions or tissues. For binary segmentation tasks, let $I$ represent the entire set of image pixels, $P$ denote the set of pixels predicted as foreground and $G$ indicate the set of actual foreground pixels based on ground truth. The detailed definition of regions is shown in Figure~\ref{fig:regions} and various evaluation metrics can be defined as follows.

\subsubsection{Global performance}

Accuracy (Acc) measures the proportion of correctly predicted pixels (both foreground and background) out of all pixels in the image.
\begin{equation}
	\mathrm{Accuracy} = \frac{|P \cap G| + |(I - P) \cap (I - G)|}{|I|}
\end{equation}
Here, $P \cap G$ and $(I - P) \cap (I - G)$ represent the set of pixels correctly predicted as foreground and background, respectively.

\subsubsection{Class-specific metrics}

For the task of binary segmentation, the term precision (Prec) is used to quantify the fraction of correctly predicted foreground pixels out of all the predicted foreground pixels. This metric is also referred to as the positive predictive value, or PPV.
\begin{equation}
	\mathrm{Precision} = \frac{|P \cap G|}{|P|}
\end{equation}

While recall (Rec) refers to the proportion of true foreground pixels that were correctly predicted as foreground.
\begin{equation}
	\mathrm{Recall} = \mathrm{Sensitivity} = \frac{|P \cap G|}{|G|}
\end{equation}

\subsubsection{Similarity and overlap}

There are two widely used metrics to evaluate the similarity between the predicted and true foreground regions: Dice similarity coefficient (DSC) and Intersection over Union (IoU). DSC, also known as the Sørensen-Dice coefficient, Dice score, or F1 score, is defined as the harmonic mean of precision and recall:
\begin{equation}
	\mathrm{DSC} = 2 \cdot \frac{\mathrm{Precision} \cdot \mathrm{Recall}}{\mathrm{Precision} + \mathrm{Recall}} = 2 \cdot \frac{|P \cap G|}{|P| + |G|}
\end{equation}

The DSC ranges from 0 to 1, with a higher value indicating greater similarity between the predicted region ($P$) and the ground truth ($G$); a value of 0 indicates no overlap, while 1 represents a perfect overlap.

IoU, also known as the Jaccard index, is defined as:
\begin{equation}
	\mathrm{IoU} = \frac{|P \cap G|}{|P \cup G|}
\end{equation}

IoU also ranges from 0 to 1, where a higher value indicates a greater degree of overlap between $P$ and $G$. Unlike DSC, which balances precision and recall, IoU provides a direct measure of the proportion of the overlapping area relative to the total area encompassed by both the prediction and the ground truth. This difference often results in slightly different values for the same segmentation performance, with IoU generally being lower than DSC when there is a partial overlap.

\subsubsection{Regional variability}

Volumetric overlap error (VOE) is a metric used to quantify the discrepancy between two volumes, typically the predicted mask and the ground truth.
\begin{equation}
	\mathrm{VOE} = 1 - \mathrm{IoU} = 1 - \frac{|P \cap G|}{|P \cup G|}
\end{equation}

As shown above, the value of VOE ranges from $0$ to $1$. A lower VOE indicates a higher overlap and thus, a more accurate segmentation.

The Hausdorff distance (HD) is an indicator of the distance between the farthest points of two sets of points, used to assess the similarity between two shapes, especially when considering the match of their boundaries. For the sets of foreground pixels, $P$ and $G$, the Hausdorff distance can be defined as:
\begin{equation}
	\mathrm{HD} = \max \left\{\sup _{p \in P} \inf _{g \in G} d(p, g), \sup _{g \in G} \inf _{p \in P} d(g, p)\right\}
\end{equation}
where $\sup$ indicates the supremum operator, $\inf$ refers the infimum operator, and $d(p, g)$ represents the Euclidean distance between pixels $p$ and $g$ of the set $P$ and $G$, respectively.

In order to eliminate the effect of outliers on HD, the 95th percentile HD is commonly used as an evaluation metric to enhance the robustness of shape assessment, which is also known as HD95. The smaller the HD, the smaller the maximum deviation between $P$ and $G$.

Relative volume difference (RVD) quantifies the relative difference in volume (number of pixels) between the predicted and actual foreground.
\begin{equation}
	\mathrm{RVD} = \frac{|P| - |G|}{|G|}
\end{equation}

A positive RVD value indicates that the predicted volume is larger than the actual volume, and vice versa. A smaller absolute RVD value means a smaller difference in volume between $P$ and $G$.

Average symmetric surface distance (ASD) measures the average distance between the boundaries of the predicted and actual foreground, symmetrically considering the distances in both directions.
\begin{equation}
	\mathrm{ASD} = \frac{\sum_{p \in S_P} \min _{g \in S_G} d(p, g)+\sum_{g \in S_G} \min _{p \in S_P} d(g, p)}{\left|S_P\right|+\left|S_G\right|}
\end{equation}
where $S_P$ and $S_G$ are the sets of boundary pixels for the predicted and actual foregrounds, respectively.

Similar to HD, the ASD value is greater than or equal to $0$, where a smaller ASD value represents a smaller average distance between $P$ and $G$.

\subsubsection{Summary}

The detailed calculation process of evaluation metrics used in the included studies are summarized in Table~\ref{tab:summary_of_metrics}. The value range, unit, size preference, and expression of each metric are provided in the table.

\renewcommand{\arraystretch}{1.5}

\begin{table*}[htb]
	\caption{Summary of evaluation metrics. For unit, / refers to dimensionless. For size preference, $\uparrow$ indicates the bigger value is better, and vice versa. The $I$, $P$ and $G$ represent the base image, predicted and ground truth binary masks, respectively. The $S_P$ and $S_G$ represent the set of boundary pixels in $P$ and $G$, respectively, and $d(p, g)$ represents the Euclidean distance between point $p$ and $g$.}
	\centering
	\footnotesize
	\setlength{\tabcolsep}{0.5em}
	\begin{tabular}{lllllll}
		\toprule
		Type                                                   & Metric & Value Range          & Unit & \makecell[l]{Size\\Preference} & Formula                                                                                                                           \\
		\midrule
		\makecell[l]{Global\\performance}                      & Acc    & $[0, 1]$             & /    & $\uparrow$                     & $\frac{|P \cap G| + |(I - P) \cap (I - G)|}{|I|}$                                                                                 \\\hline
		\multirow{2}{*}{Class-specific}                        & Prec   & $[0, 1]$             & /    & $\uparrow$                     & $\frac{|P \cap G|}{|P|}$                                                                                                          \\
		                                                       & Rec    & $[0, 1]$             & /    & $\uparrow$                     & $\frac{|P \cap G|}{|G|}$                                                                                                          \\\hline
		\multirow{2}{*}{\makecell[l]{Overlap and\\similarity}} & DSC    & $[0, 1]$             & /    & $\uparrow$                     & $2 \cdot \frac{|P \cap G|}{|P| + |G|}$                                                                                            \\
		                                                       & IoU    & $[0, 1]$             & /    & $\uparrow$                     & $\frac{|P \cap G|}{|P \cup G|}$                                                                                                   \\\hline
		\multirow{4}{*}{\makecell[l]{Regional\\variability}}   & VOE    & $[0, 1]$             & /    & $\downarrow$                   & $1 - \frac{|P \cap G|}{|P \cup G|}$                                                                                               \\
		                                                       & HD     & $[0, +\infty)$       & mm   & $\downarrow$                   & $\max \left\{\sup _{p \in P} \inf _{g \in G} d(p, g), \sup _{g \in G} \inf _{p \in P} d(g, p)\right\}$                            \\
		                                                       & RVD    & $(-\infty, +\infty)$ & /    & Optimal at 0                   & $\frac{|P| - |G|}{|G|}$                                                                                                           \\
		                                                       & ASD    & $[0, +\infty)$       & /    & $\downarrow$                   & $\frac{\sum_{p \in S_P} \min _{g \in S_G} d(p, g)+\sum_{g \in S_G} \min _{p \in S_P} d(g, p)}{\left|S_P\right|+\left|S_G\right|}$ \\
		\bottomrule
	\end{tabular}
	\label{tab:summary_of_metrics}
\end{table*}

\renewcommand{\arraystretch}{1.0}

\section{Results}\label{sec:result}

Deep learning techniques have made significant advancements in computer vision tasks. Compared to traditional digital image processing methods, deep learning provide higher efficiency, higher accuracy, and more generalizable solutions. Deep learning models are also able to perform some difficult work in terms of traditional digital image processing, such as image generation, image super-resolution, and video frame interpolation. In the following subsections, the recent applications and developments of deep learning methods and techniques for LN segmentation, and detailed information will be discussed.

\subsection{Convolutional neural networks}

\begin{figure}[htb]
	\centering
	\includegraphics[width=0.7\linewidth]{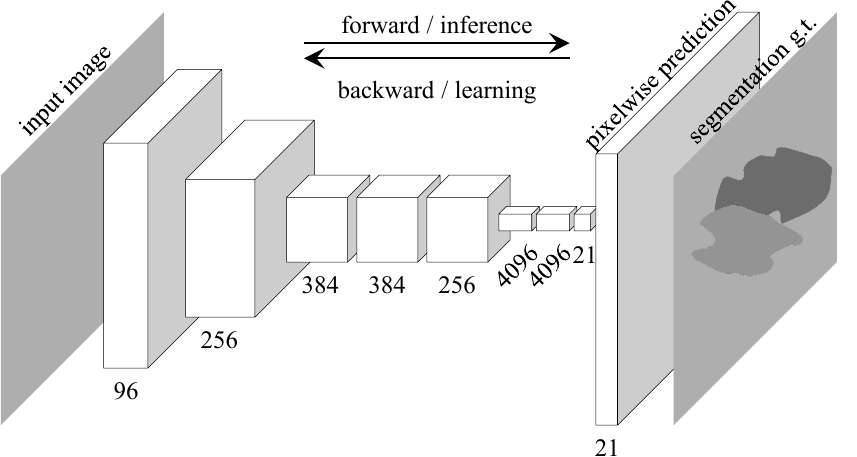}
	\caption{Overview of FCN for semantic segmentation~\cite{fcn_long_2015}.}
	\label{fig:FCN}
\end{figure}

Convolutional neural networks (CNN) have been widely used in computer vision since the introduction of LeNet~\cite{lenet_lecun_1998} and AlexNet~\cite{alexnet_krizhevsky_2017}. A CNN model typically consists of an input layer, multiple hidden layers, and an output layer, where the hidden layers commonly include convolutional layers, pooling layers, activation layers, and fully connected layers (also known as FC layers, or multi-layer perceptron, MLP). The CNN architecture is called fully convolutional networks (FCN)~\cite{fcn_long_2015} if the entire CNN architecture does not contain any FC layers. Since it is not necessary to specify the neurons explicitly, FCN can process input images of arbitrary resolution. The original FCN model proposed in 2015 for semantic segmentation is shown in Figure~\ref{fig:FCN}.

Nogues~\etal~\cite{holistically_nogues_2016} proposed a novel method for automatic segmentation of axillary LN clusters in CT images, which is a variation of FCN that combines holistically-nested neural networks (HNN)~\cite{hnn_xie_2015} and structured optimization. The area information and boundary discontinuities of the LN clusters are learned by HNN-A and HNN-C models, respectively. The experiments were conducted on a public LN dataset which contains 173 3D CT scans and yields 39,361 images after data augmentation. The substantial size of the dataset allowed for comprehensive training of the HNN, enhancing the ability of proposed model to generalize across diverse anatomical variations. Additionally, the high quality of the CT images (512$\times$512$\times$512 voxels) ensured that the edge information and boundary discontinuities of the LN clusters were accurately captured, which is critical for the effectiveness of the boundary neural fields structured optimization. This combination of a large and high-quality dataset contributed significantly to the superior accuracy achieved by the HNN method compared to other optimization techniques in LN group volume measurements, demonstrating the value of edge detection methods for LN segmentation tasks.

\begin{figure}[htb]
	\centering
	\subfloat[The architecture of FCN module]{
		\centering
		\includegraphics[width=0.8\linewidth]{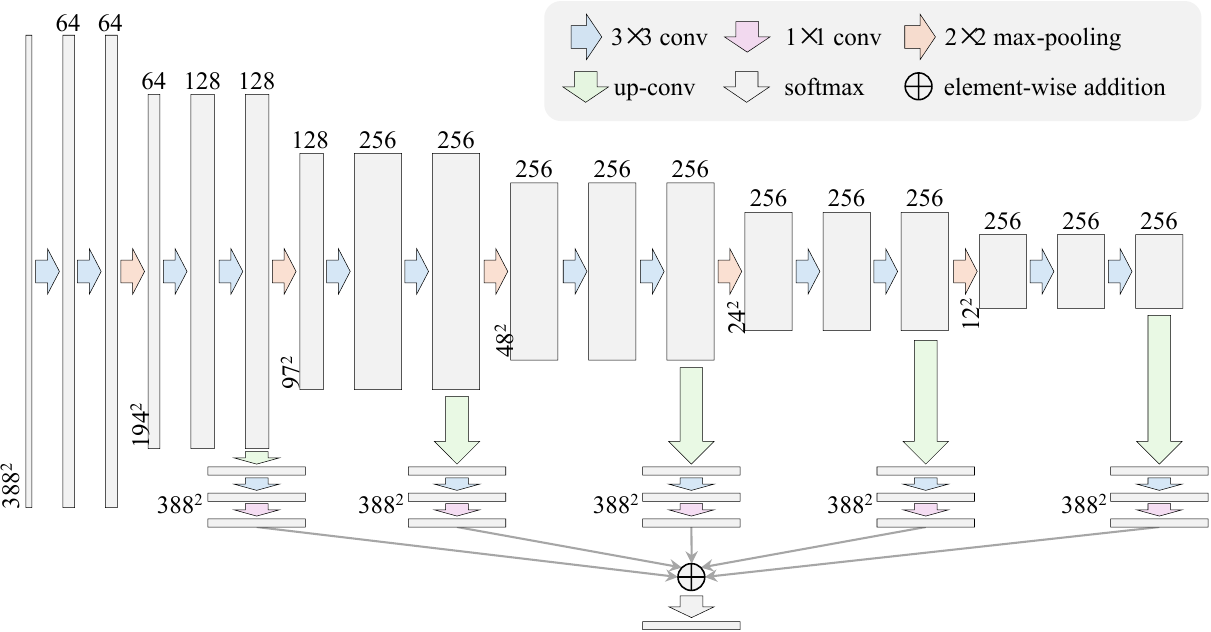}
	}
	\\ \vspace{0.5em}
	\subfloat[Workflow of CFS-FCN]{
		\centering
		\includegraphics[width=0.7\linewidth]{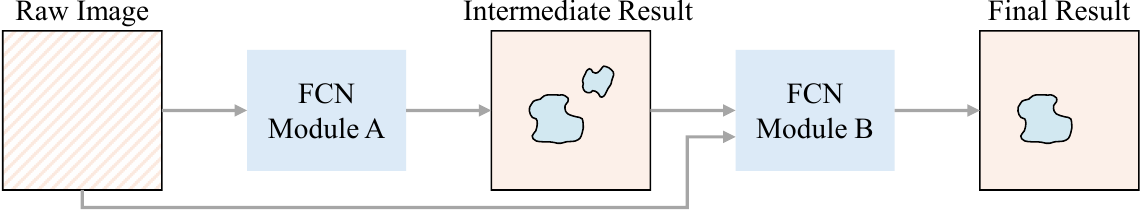}
	}
	\caption{Overview of CFS-FCN~\cite{coarse_zhang_2016}.}
	\label{fig:CFS-FCN}
\end{figure}

Another novel deep learning approach based on FCN for the automatic segmentation of LNs in ultrasound images named coarse-to-fine stacked fully convolutional nets (CFS-FCN) was introduced by Zhang~\etal~\cite{coarse_zhang_2016}. The CFS-FCN model consists of two parts: (1) using the first FCN model to generate intermediate coarse masks of LNs, then (2) combining coarse masks with original input for the second FCN model to generate the final fine mask of LNs. The overview architecture of CFS-FCN is illustrated in Figure~\ref{fig:CFS-FCN}. The dataset used in this study consists of 80 images, and the results show that the CFS-FCN model with boundary refinement technology significantly outperforms existing deep learning approaches, even with a small amount of data.

Zhang~\etal~\cite{decompose_zhang_2019} presented a new generalizable strategy for medical image segmentation, named decompose-and-integrate learning. It divides the segmentation task into sub-problems (decomposition phase) solved by deep learning modules, each with unique feature transformations. These solutions are then combined (integration phase) to solve the original segmentation problem. This method was evaluated on model DenseVoxNet~\cite{densevoxnet_yu_2017} and CUMedNet~\cite{cumednet_chen_2016} in 3D and 2D  images, respectively. The ablation experiments conducted on multiple datasets (including an in-house ultrasound dataset of LN) demonstrate the robustness and generalization capability of different data of the proposed strategy.

\begin{figure}[htb]
	\centering
	\includegraphics[width=0.8\linewidth]{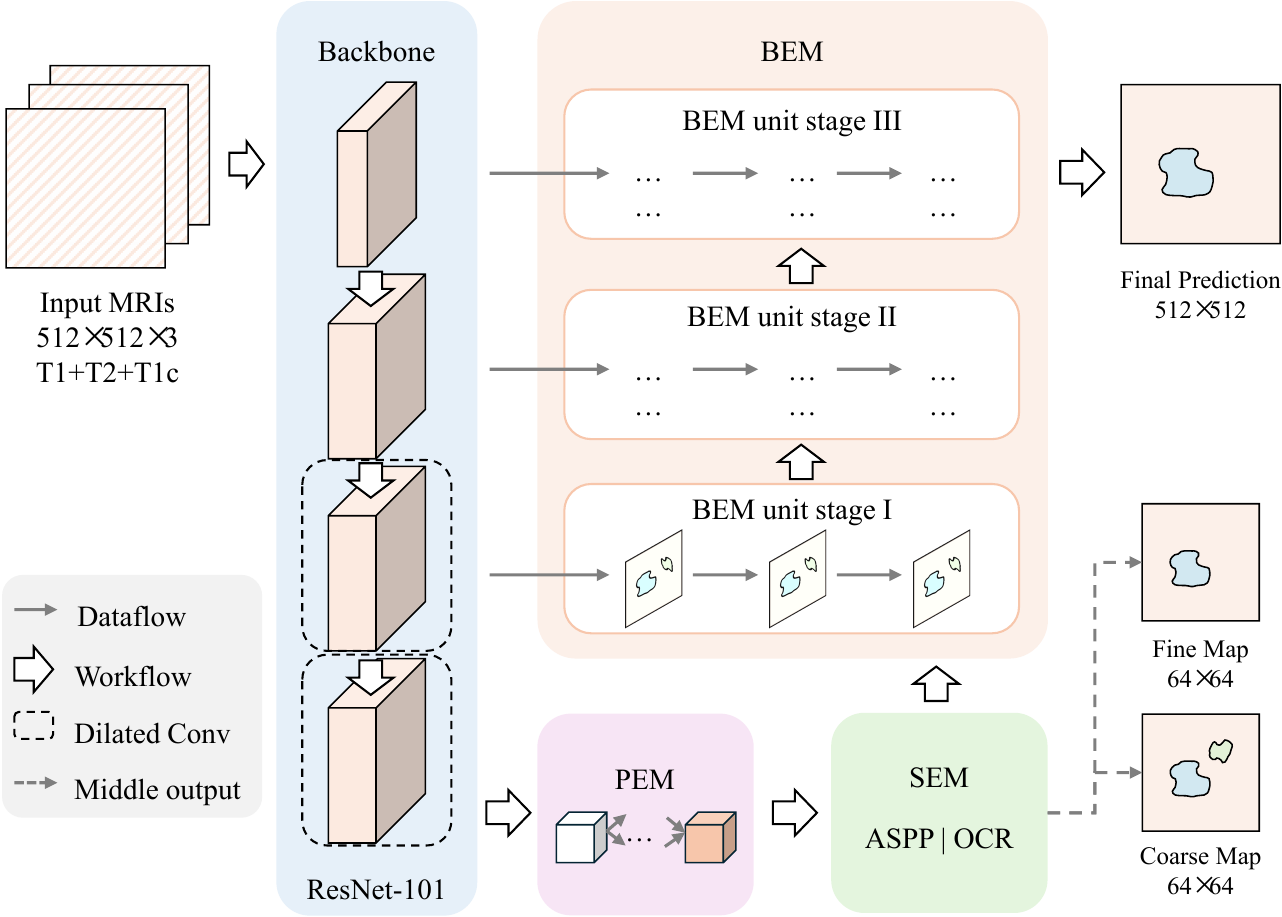}
	\caption{Overview of NPCNet~\cite{npcnet_li_2022}.}
	\label{fig:NPCNet}
\end{figure}

In addition to being used in CT and ultrasound image segmentation tasks, FCN has also been used in MRI image segmentation tasks. Li~\etal~\cite{npcnet_li_2022} introduced NPCNet for the precise segmentation of primary nasopharyngeal carcinoma (NPC) tumors and metastatic LNs (MLNs) in MRI images. The NPCNet model incorporates three key modules: position enhancement module (PEM), scale enhancement module (SEM), and boundary enhancement module (BEM), which is similar to Zhang~\etal~\cite{coarse_zhang_2016}, and aimed to address the challenges related to variable localization, variable size, and irregular boundaries of MLNs. The structure of NPCNet is illustrated in Figure~\ref{fig:NPCNet}. Notably, NPCNet adopted the pre-trained (on ImageNet~\cite{imagenet_deng_2009}) ResNet-101 as the backbone. Through extensive experimentation on a dataset comprising 9,124 samples from 754 patients, the model demonstrated state-of-the-art performance in segmenting NPC tumors and MLNs, outperforming other popular segmentation models.

In summary, considerable research has been dedicated to the application of CNN and FCN models in the segmentation of LNs in CT, PET/CT, ultrasound, and MRI images. The FCN model has also been extensively utilized in LN segmentation tasks due to its capacity to process images of arbitrary resolution and its ability to effectively capture spatial information. Although CNN models demonstrate efficient local feature extraction capabilities and a well-established technological foundation in LN segmentation tasks, they are deficient in multi-scale feature extraction and global context modeling. Consequently, there has been an increased focus on enhancing CNN model performance by incorporating additional modules and structures, including the encoder-decoder structure with skip connections, Transformer architecture based on the attention mechanism, and advanced loss function design.

\subsection{Encoder-Decoder networks}

\begin{figure}[htb]
	\centering
	\includegraphics[width=0.9\linewidth]{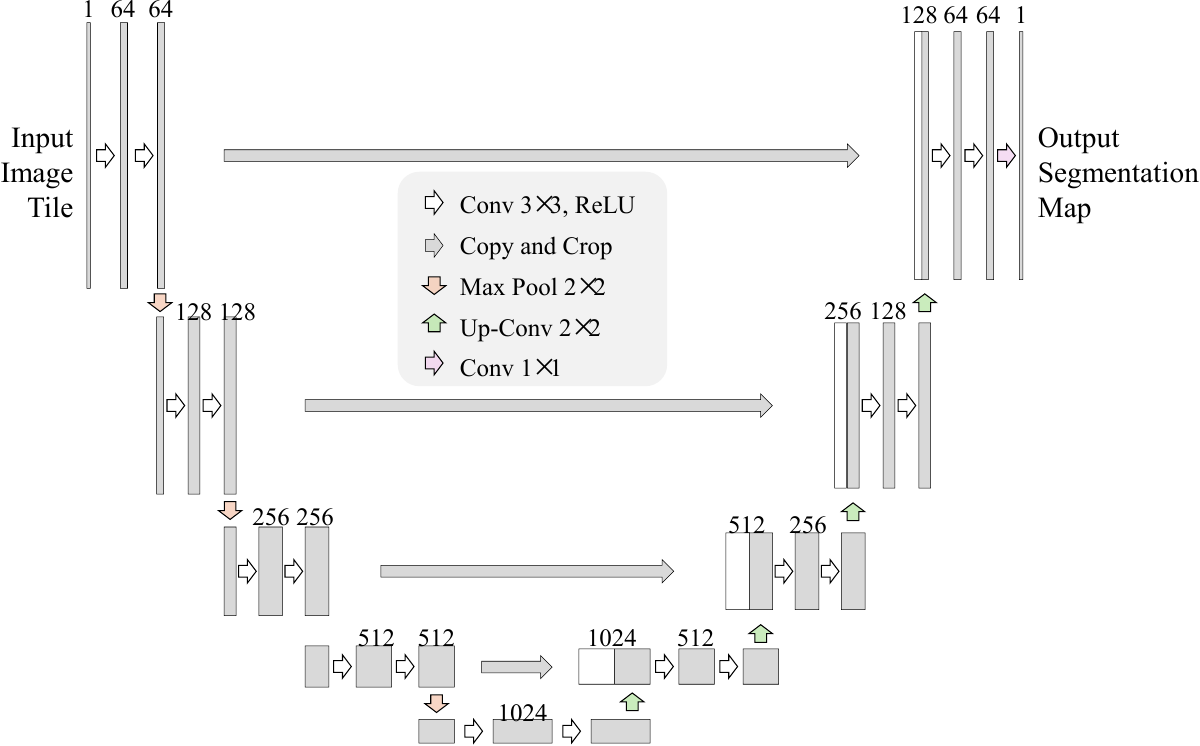}
	\caption{Overview of U-Net~\cite{unet_ronneberger_2015}.}
	\label{fig:U-Net}
\end{figure}

Since the introduction of U-Net~\cite{unet_ronneberger_2015}, the encoder-decoder structure has quickly become the standard choice for medical image segmentation. This is due to the advantages it offers over other approaches, including a lightweight network structure, multi-scale feature extraction mechanism, and the preservation of spatial information through the skip connection design. As shown in Figure~\ref{fig:U-Net}, U-Net gets its name because of a unique U-shaped structure. It consists of two parts: a contraction path on the left side (encoder) and a symmetric expansion path on the right side (decoder). The encoder part extracts image features and reduces their dimensionality through successive convolution and pooling operations, while the decoder part gradually recovers the spatial resolution and detailed information of the image using up-sampling. The key feature of the U-Net is the introduction of skip connections between the encoder and decoder at corresponding levels, effectively preserving rich contextual information and significantly enhancing the accuracy of image segmentation.

The encoder-decoder structure has been widely used in the segmentation of LNs in CT and PET/CT images. Men~\etal~\cite{ddnn_men_2017} proposed an end-to-end deep deconvolutional neural network (DDNN) based on the encoder-decoder structure, and aimed at accelerating the segmentation process of NPC target areas in CT scan images for radiotherapy. Utilizing data from 230 patients for the training and testing process, the study demonstrated that the DDNN outperforms the VGG-16~\cite{vgg_simonyan_2014} model across all segmentation tasks. However, the segmentation accuracy of DDNN for LN gross tumor volume was relatively lower due to variations in shape, volume, and location among patients.

Similarly, Li~\etal~\cite{tumor_li_2019} have also investigated the segmentation of NPC LNs in CT scan images. They proposed a modified U-Net model to improve the segmentation accuracy of NPC LNs, which generates segmentation results with the same resolution as the input image. This study mainly focuses on the diagnosis of different stages of NPC primary tumors and metastatic LNs based on the same deep learning model. The experimental results showed that the proposed model achieves a slightly higher segmentation accuracy for LNs at stage N1 (0.691) than stages N2 (0.653) and N3 (0.640), while N2 and N3 represent more advanced stages of cancer with more complex anatomical changes, leading to lower segmentation accuracy.

Ariji~\etal~\cite{seg_ariji_2022} utilized U-Net to automatically segment multiple classes of cervical LNs from enhanced CT images of patients with oral cancer and classify metastatic or non-metastatic LNs. Nayan~\etal~\cite{mediastinal_nayan_2022} proposed an enhanced UNet++~\cite{unet++_zhou_2018} model to achieve high-precision automatic detection and segmentation of mediastinal LNs from CT images. It is worth noting that in this study, bilinear interpolation was used instead of transposed convolution for upsampling operations in the decoder path of UNet++. This approach was used to reduce computational intensity and avoid the introduction of checkerboard artifacts that are commonly associated with transposed convolution. The complete dataset used for experiments consisted of three separate datasets, including 54,330 images after data augmentation. The results showed that the enhanced UNet++ model achieved superior performance in mediastinal LN detection and segmentation tasks, outperforming the original UNet++ model and other advanced methods.

\begin{figure}[htb]
	\centering
	\includegraphics[width=0.8\linewidth]{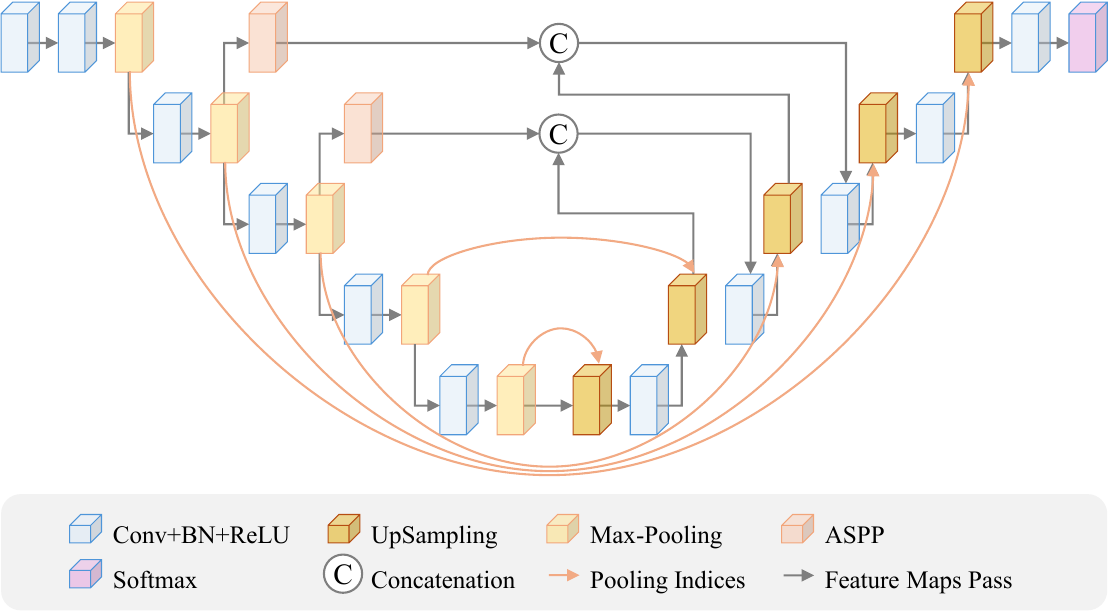}
	\caption{Overview of DiSegNet~\cite{disegnet_xu_2021}.}
	\label{fig:DiSegNet}
\end{figure}

Some researchers have also attempted to combine PET images with CT images to perform LN segmentation tasks. Xu~\etal~\cite{disegnet_xu_2021} proposed DiSegNet for LN segmentation in PET/CT images. This study included a new cosine-sine (CS) loss function to address the class imbalance problem for different networks during training and the incorporation of a multi-stage atrous spatial pyramid pooling (MS-ASPP) submodule to leverage multi-scale information for enhanced segmentation performance of LN boundaries. The overview structure of DiSegNet is shown in Figure~\ref{fig:DiSegNet}. The DiSegNet architecture enhances the SegNet~\cite{segnet_badrinarayanan_2017} framework with the MS-ASPP module to achieve superior semantic accuracy and detail preservation in segmentation, and the encoder module of DiSegNet can be replaced with other pre-trained models such as ResNet to improve the segmentation performance.

Ahamed~\etal~\cite{seg_ahamed_2023} developed an automated segmentation approach based on the U-Net architecture with a ResNet50 encoder pre-trained on ImageNet, aimed at segmenting primary tumors and metastatic LNs from PET/CT images of head and neck cancer patients. Similar to Xu~\etal~\cite{disegnet_xu_2021}, this study proposed a multiclass Dice loss function combining primary tumor and LN segmentation loss to optimize model training. The encoder-decoder structure is asymmetrical, with the decoder path being shallower than the encoder (approximately 4:1 ratio). Results demonstrate the potential of this asymmetric structure in improving the efficiency and accuracy of medical image analysis.

\begin{figure}[htb]
	\centering
	\subfloat[The architecture of FSM]{
		\centering
		\includegraphics[width=0.6\linewidth]{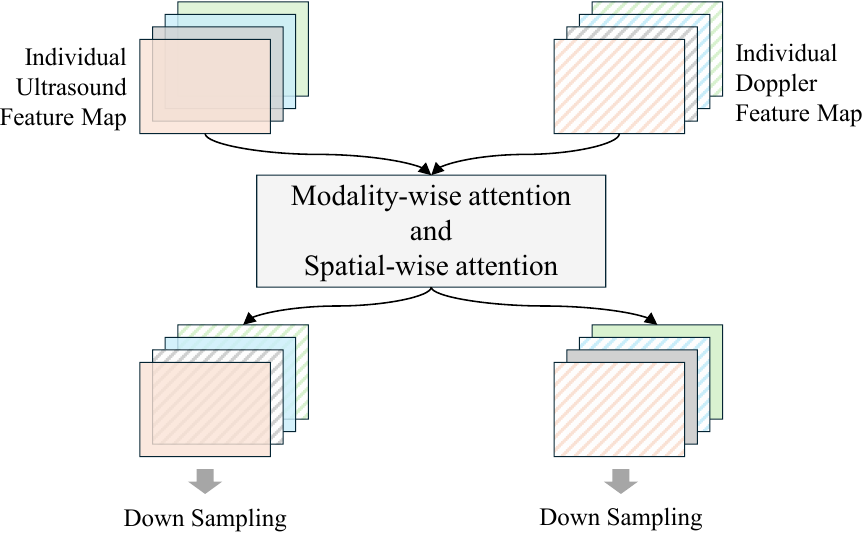}
	}
	\\
	\subfloat[Modality-wise attention]{
		\centering
		\includegraphics[width=0.45\linewidth]{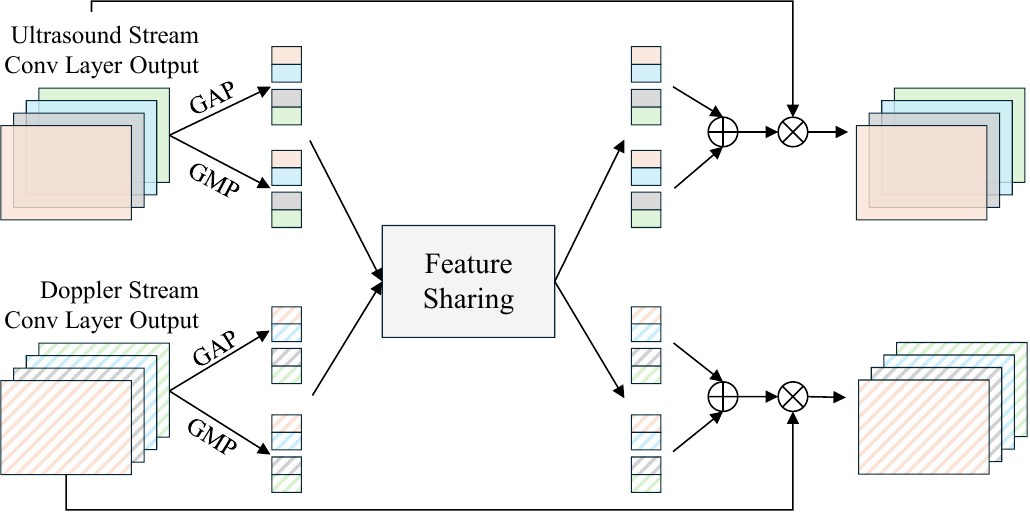}
	}
	\subfloat[Spatial-wise attention]{
		\centering
		\includegraphics[width=0.45\linewidth]{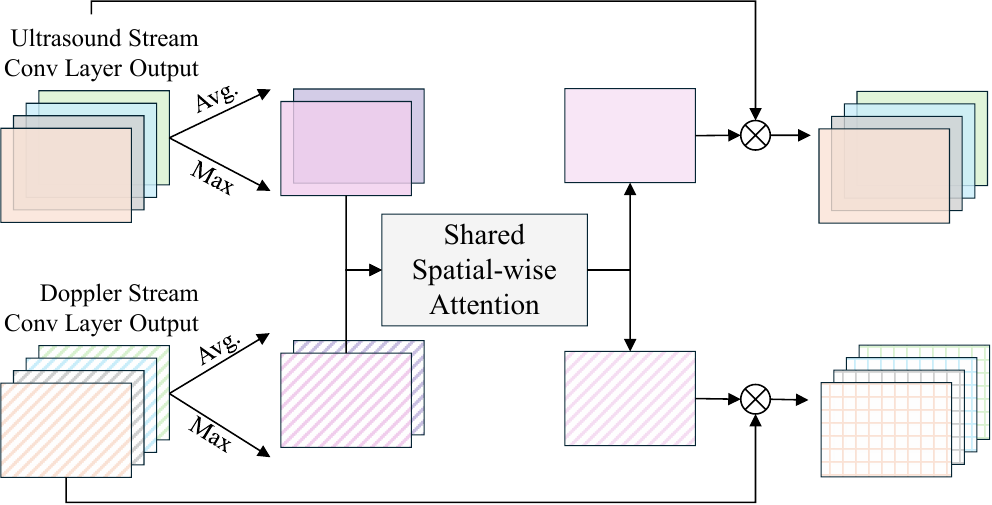}
	}
	\caption{Overview of multi-modal feature attention~\cite{multi_modal_fu_2020}.}
	\label{fig:multi_modal}
\end{figure}

The encoder-decoder structure has also been applied to LNs segmentation in ultrasound images. Fu~\etal~\cite{multi_modal_fu_2020} presented a multimodal fusion method for the cervical LNs segmentation from fused features of grayscale and Doppler ultrasound images. The core design is the feature attention mechanism that utilizes the information of higher dimensions provided by both imaging modalities. Unlike the attention mechanism in the Transformer~\cite{transformer_vaswani_2017}, this feature attention mechanism is designed to exchange and fuse the modality-wise and spatial-wise features. The proposed feature-sharing module (FSM) suppresses the irrelevant features while highlighting the key features required to differentiate the LNs from the surrounding tissues. The FSM, modality-wise attention, and spatial-wise attention are illustrated in Figure~\ref{fig:multi_modal}. This study also adopted several pre-processing techniques to enhance the quality of ultrasound images, such as auto cropping based on single shot multibox detector (SSD)~\cite{ssd_liu_2016}, noise reduction, and modalities registration. This multimodal fusion method significantly improved LN segmentation accuracy, outperformed the coarse-to-fine method proposed by Zhang~\etal~\cite{coarse_zhang_2016} and U-Net, and also marked an important advancement in the application of multimodal data fusion for medical image processing.

\begin{figure}[htb]
	\centering
	\includegraphics[width=0.8\linewidth]{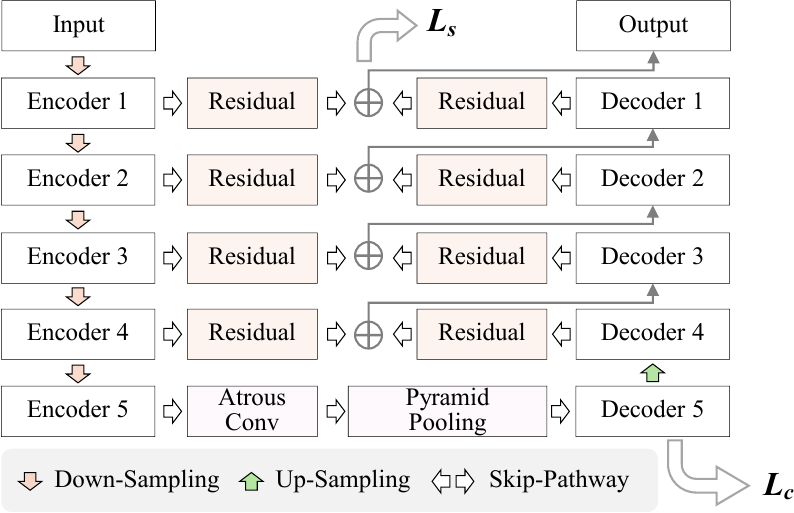}
	\caption{Overview of MA-Net~\cite{manet_zhang_2020}.}
	\label{fig:MA-Net}
\end{figure}

Zhang~\etal~\cite{manet_zhang_2020} introduced MA-Net, a multi-attention and atrous convolution network designed to enhance semantic information extraction through an end-to-end approach. As shown in Figure~\ref{fig:MA-Net}, the network is based on an encoder-decoder architecture that combines multiple-channel convolution blocks, atrous convolution modules, pyramid pooling modules, residual skip connections, and a multi-task loss function (composed of cross-entropy loss $L_c$ and Dice loss $L_s$ in a certain ratio). Segmentation experiments were conducted on multiple datasets including ultrasound images of the brachial plexus, fetal head, and LN. The results showed that MA-Net achieves significant improvements in mainstream performance evaluation metrics compared to U-Net and UNet++~\cite{unet++_zhou_2018}. It it highly generalizable and practical for accurate ultrasound, MRI and CT image segmentation.

\begin{figure}[htb]
	\centering
	\includegraphics[width=0.9\linewidth]{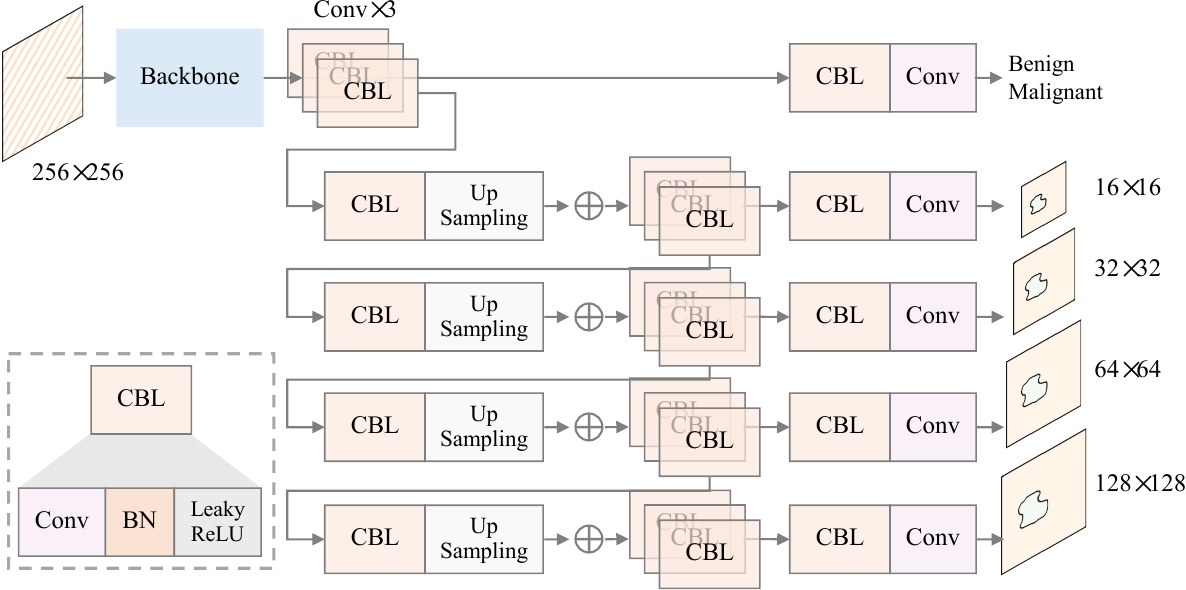}
	\caption{Overview of MUNet~\cite{munet_zhang_2020}.}
	\label{fig:MUNet}
\end{figure}

Zhang~\etal~\cite{munet_zhang_2020} introduced a multi-scale U-Net (MUNet) for ultrasound image segmentation, combining FCN, encoder-decoder architecture with a feature pyramid. The overview structure of MUNet is shown in Figure~\ref{fig:MUNet}. Similar to YOLOv3~\cite{yolov3_redmon_2018}, this model is also composed of a replaceable backbone and multi-scale segmentation branches, but with an additional branch for cervical LN classification, and it is capable of processing images at arbitrary sizes thanks to the FCN structure. The experiments were based on a dataset consists of 4,000 benign and 1,000 malignant LN images before augmentation, with a resolution of 700$\times$800, and yielded a high Dice score and AUC value. The special design of MUNet allows the backbone network to be replaced accommodating different needs for efficiency, accuracy, and different scenarios.

In addition to focusing on new methods, some researchers have also explored different ultrasound image preprocessing techniques. To overcome the challenges of speckle noise and echogenic hila that existed in ultrasound LN images, Chen~\etal~\cite{diffusion_chen_2021} proposed a method that integrates anisotropic diffusion denoising based on Gabor-based anisotropic diffusion, a modified U-Net, and morphological operations. It reveals the potential of combining traditional image processing techniques with deep learning methods to improve segmentation accuracy.

\begin{figure}[htb]
	\centering
	\includegraphics[width=0.8\linewidth]{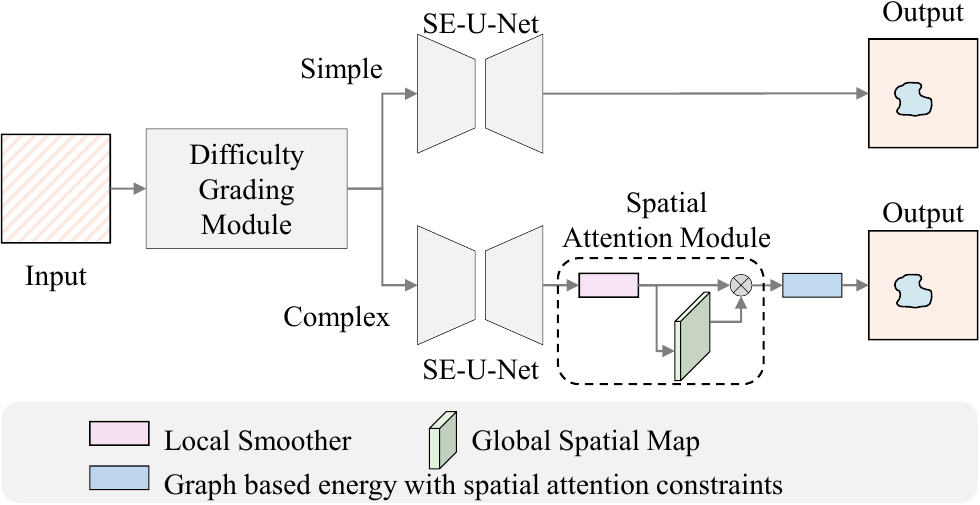}
	\caption{Overview of difficulty-aware dual network structure~\cite{difficulty_xu_2022}.}
	\label{fig:difficulty-aware}
\end{figure}

Xu~\etal~\cite{difficulty_xu_2022} proposed a difficulty-aware dual network structure for axillary LN segmentation in ultrasound images, combined with a spatial attention-constrained graph model. First, the difficulty level of the input image is evaluated by a grading module, and then different network branches are used for adaptive segmentation according to the difficulty level. The overview structure is shown in Figure~\ref{fig:difficulty-aware}.

The complex LN images refer to images with unclear LN boundaries, low contrast, and complicated intensity distribution. For dealing with those images, this study introduced the spatial attention module and a graph-based energy model that considers the constraints of spatial attention and intends to provide additional discriminative information and enhance segmentation performance by capturing inter-pixel relationships. However, the experimental results were evaluated quantitatively based on a combination of complex and simple LN images, and it lacks performance analysis specifically for complex or simple LN images. Nevertheless, the overall results show that the method outperforms other state-of-the-art deep learning methods in segmenting axillary LNs, such as U-Net, FCN-8s~\cite{fcn_long_2015}, DeepLabv3+~\cite{deeplabv3+_chen_2018}, SegNet, and Frrn~\cite{frrn_pohlen_2017}.

\begin{figure}[htb]
	\centering
	\includegraphics[width=0.8\linewidth]{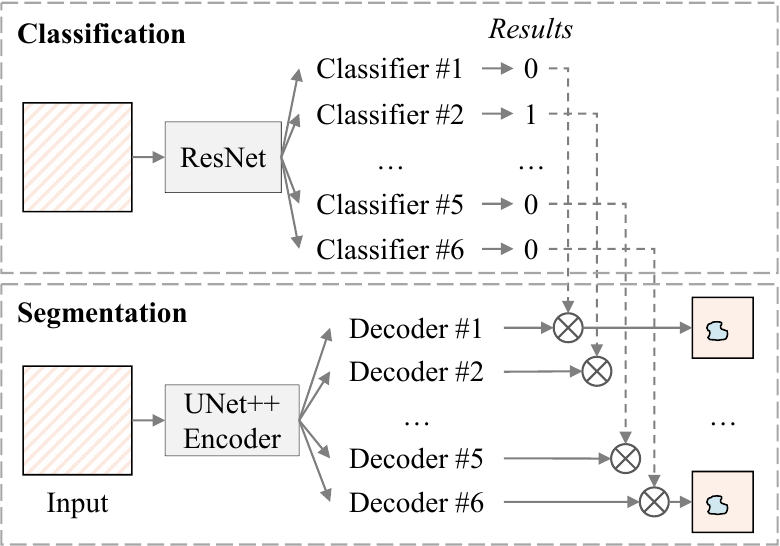}
	\caption{Overview of CMU-Net~\cite{efficient_wen_2024}.}
	\label{fig:cmu_net}
\end{figure}

Wen~\etal~\cite{efficient_wen_2024} concentrated on the segmentation of six LN regions (LNRs) in CT images of patients with rectal, prostate, and cervical cancer, including abdominal presacral, pelvic presacral, internal iliac nodes, external iliac nodes, obturator nodes, and inguinal nodes. A cascaded multi-heads U-net (CMU-net) was proposed to classify and segment the six LNRs simultaneously. The classification model was constructed using the ResNet-50~\cite{resnet_kaiming_2016}, while and the segmentation model was based on the UNet++~\cite{unet++_zhou_2018}. Six distinct heads were employed to predict the six LNRs, and the final segmentation results were obtained by combining the six segmentation results correspondingly. The overview structure of CMU-Net is shown in Figure~\ref{fig:cmu_net}.

The classification and segmentation model were trained and validated on 120 cases, another 40 cases were used for testing. All images were resized to 512$\times$512 pixels without augmentation. With the prior knowledge from the classification model, the performance of segmentation model achieved an average Dice score of 0.895.

\begin{figure}[htb]
	\centering
	\includegraphics[width=0.9\linewidth]{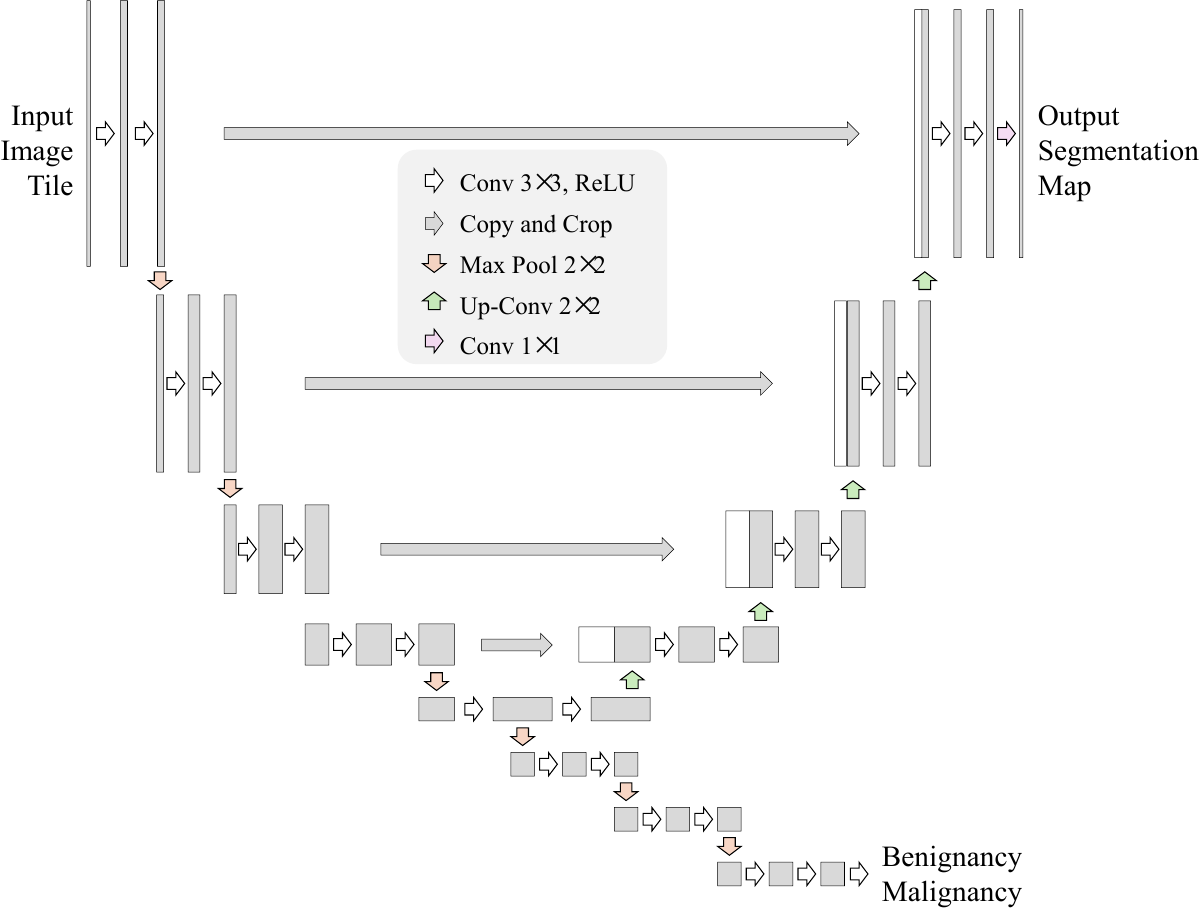}
	\caption{Overview of Y-Net~\cite{deep_zhao_2024}.}
	\label{fig:y_net}
\end{figure}

Additionally, Zhao~\etal~\cite{deep_zhao_2024} also addressed the classification and segmentation of LNs in a simultaneous manner. A novel Y-Net architectural modification was derived from the U-Net, incorporating a classification branch into the original U-Net structure. This was devised to predict the LN status (benignancy or malignancy). The overall structure of Y-Net is shown in Figure~\ref{fig:y_net}. The dataset for model training contains 2,512 images for training and testing, and 547 images for validation. The results show that two parallel branches reached 72.03\% accuracy, which outperformed the original ultrasonic report by 7.37\%. The segmentation branch obtained a median Dice score of 0.832, which is comparable to the state-of-the-art methods.

\begin{figure}[htb]
	\centering
	\subfloat[Spatial context network with atrous spatial pyramid pooling for lymph node segmentation]{
		\centering
		\includegraphics[width=0.9\linewidth]{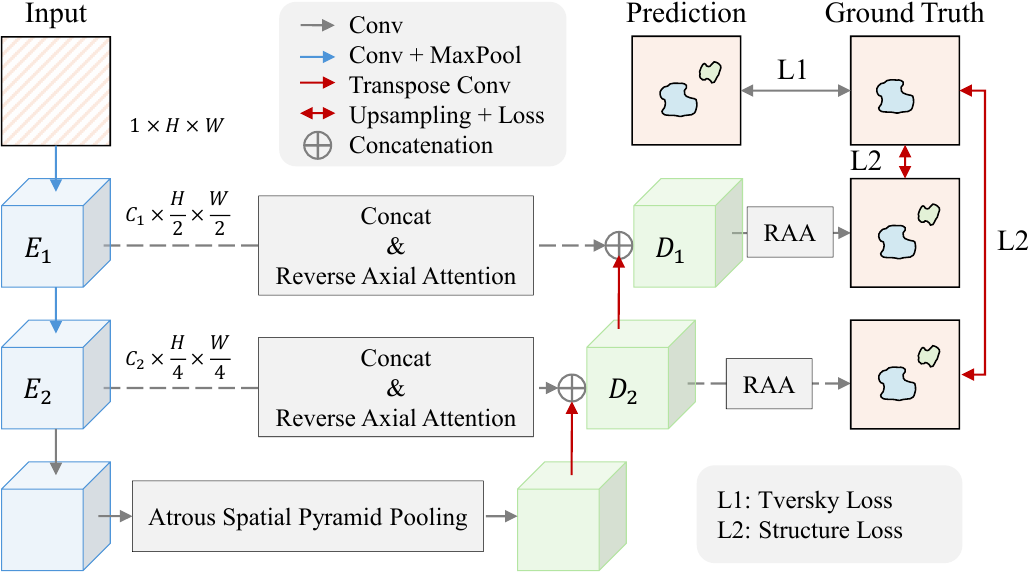}
	}
	\\ \vspace{0.5em}
	\subfloat[Reverse axial attention (RAA) in decoder layers]{
		\centering
		\includegraphics[width=0.75\linewidth]{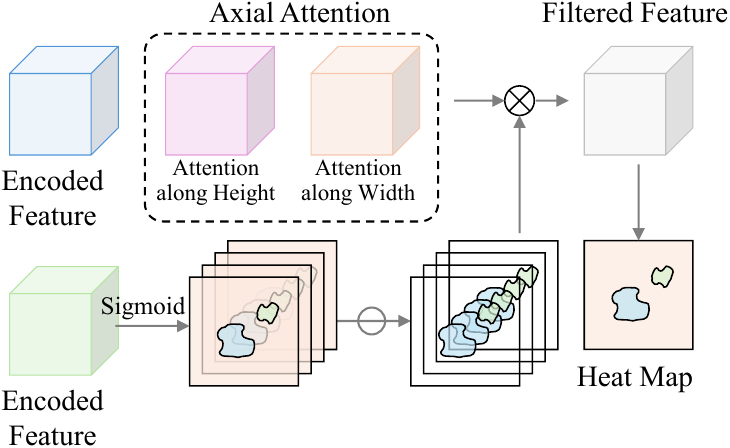}
	}
	\caption{Overview of proposed attentional U-Net for lymph node segmentation~\cite{automated_hasan_2024}.}
	\label{fig:attention_unet}
\end{figure}

In addition to the residual connections that are commonly used in encoder-decoder architectures, researchers have also considered the potential benefits of introducing attention mechanisms into UNets. Hasan~\etal~\cite{automated_hasan_2024} proposed an attentional U-Net with spatial context network and reverse axial attention for 2D LN segmentation. The spatial context network is designed to capture the spatial context information of input 2D CT slices, while the reverse axial attention mechanism can enhance the feature representation of the decoder layers. The overview of the proposed attentional U-Net is shown in Figure~\ref{fig:attention_unet}.

Interestingly, Hasan~\etal~\cite{automated_hasan_2024} focused on the segmentation of normal, small LNs in the neck of healthy individuals, which presents more challenges compared to abnormal LN segmentation tasks due to the smaller size and less distinct boundaries. They collected 221 contrast-enhanced CT scans consisting of 25,119 CT slices of the neck, with 18,054 slices used for training, 4,463 slices for validation, and 2,602 slices for testing. It achieved promising  segmentation results with a 0.808 Dice score. This study designed advanced reverse axial attention (RAA) module (composed by two 1D self-attention along height and width) and an improvement of IoU metric by 0.06 was found (from 0.774 to 0.780). The RAA module helps the model focus on relevant regions by filtering out noise and enhancing the salient features of small LNs. This mechanism is particularly effective in capturing the multi-scale context of lymph nodes, which vary in size and have irregular boundaries.

To summarize, the encoder-decoder structure has been extensively employed in the segmentation of LNs in CT, PET/CT, and ultrasound images. The U-Net~\cite{unet_ronneberger_2015} and its variants have been the most frequently utilized models by researchers in the domain of medical image segmentation. The U-Net~\cite{unet_ronneberger_2015} architecture has been modified and enhanced in various ways to improve the accuracy and efficiency of LN segmentation tasks. The introduction of attention mechanisms, feature-sharing modules, and multi-scale segmentation branches has demonstrated significant potential in enhancing the performance of deep learning models for LN segmentation. The combination of traditional image processing techniques and deep learning methods has also been investigated to improve the quality of ultrasound images and enhance the segmentation accuracy of LNs.

\subsection{Transformer}

In recent years, the Transformer~\cite{transformer_vaswani_2017,vit_dosovitskiy_2021,swint_liu_2021} architecture has been introduced into the field of image processing. By efficiently extracting complex spatial structural information and capturing global contextual relationships, the Transformer offers new possibilities for improving the accuracy and efficiency of medical image segmentation. This is particularly beneficial when dealing with high-resolution medical image data, as the Transformer can understand the overall structure of the image while preserving detailed information. As a result, it has shown great potential in tasks such as tumor identification and organ delimitation, leading to more accurate segmentation results.

\begin{figure}[htb]
	\centering
	\includegraphics[width=0.9\linewidth]{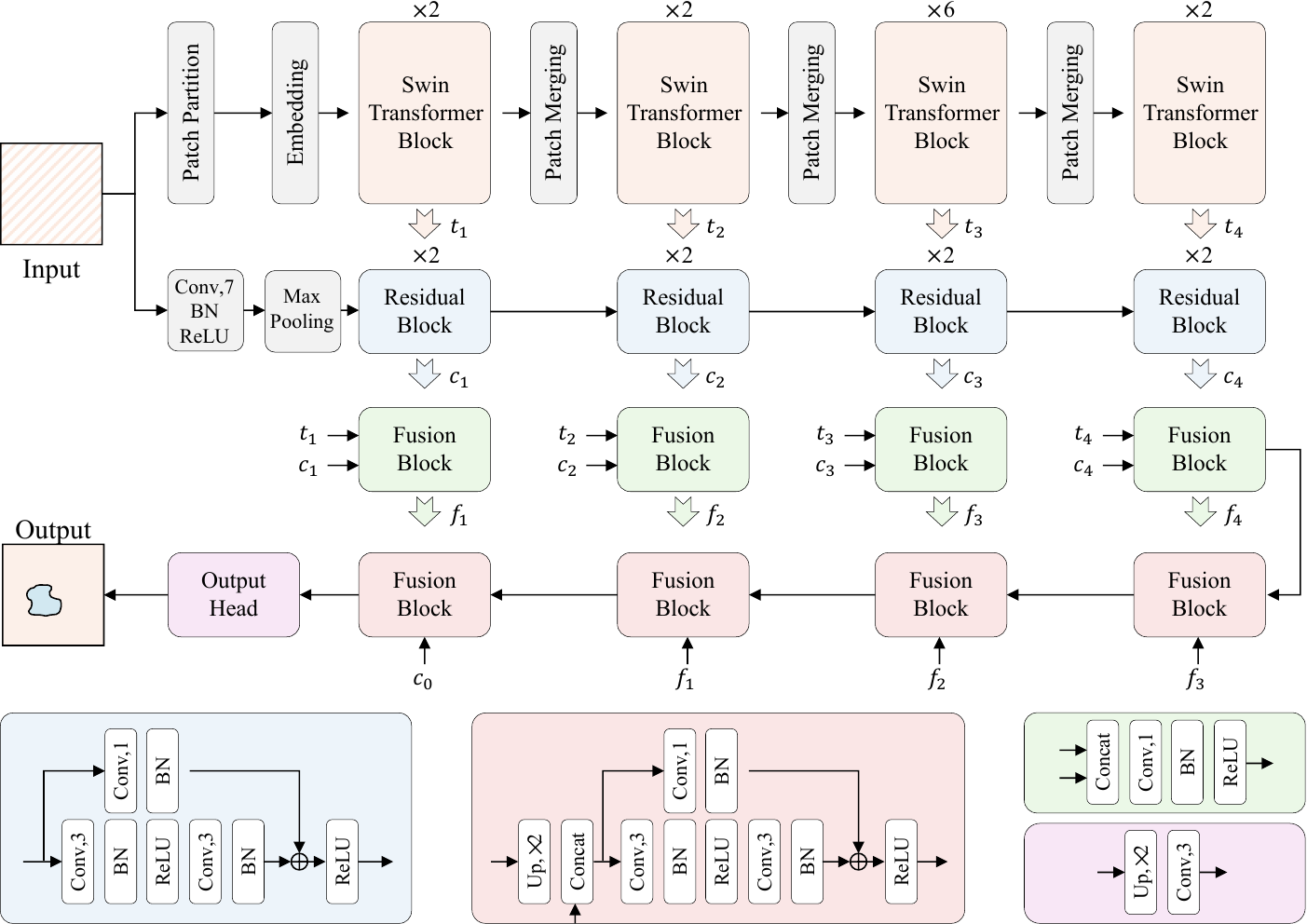}
	\caption{Overview of DE-Net~\cite{denet_shi_2022}.}
	\label{fig:DE-Net}
\end{figure}

Shi~\etal~\cite{denet_shi_2022} proposed an innovative dual-encoder hybrid model called DE-Net, which is designed to automatically segment multiple structures for whole bone marrow and lymphatic irradiation in bone marrow transplantation. The architecture of DE-Net is shown in Figure~\ref{fig:DE-Net}. The structure is still U-shaped but with two encoders and one decoder, where the first encoder is comprised of ResNet blocks and the second encoder is based on Swin Transformer blocks. DE-Net fully exploits both local detailed spatial information and global contextual knowledge in parallel to improve the quality of learned features.

Experiments were conducted on a dataset comprising seven target structures, including skulls, ribs, and LNs. The results demonstrate that DE-Net exhibits slightly enhanced performance compared to existing methods in the context of segmentation. It attained marginal improvements of 0.01 and 0.004 in the Dice score for LNs and on average, respectively, when benchmarked against another CNN and Transformer hybrid model, UTNet~\cite{utnet_gao_2021}. However, while DE-Net leverages a hybrid architecture, the broader adoption of transformer-based models in LN segmentation remains limited. This is not only due to the computational cost of transformers, but also due to other critical factors, such as the scarcity of large-scale, high quality annotated LN datasets required for effective pre-training and the inherent anatomical complexity of LNs.

Moreover, recent developments in lightweight transformer variants, including MedViT~\cite{medvit_manzari_2023}, EfficientViT~\cite{efficientvit_cai_2023}, and MobileViT~\cite{mobilevit_mehta_2022}, offer computationally efficient alternatives with comparable performance. These architectures leverage optimized attention mechanisms and parameter-efficient designs, making them more suitable for medical image segmentation, including LN segmentation. However, such lightweight models still require large-scale datasets for training and have not been extensively explored in the context of LN segmentation, highlighting an area for future research.

\subsection{Object detection assisted segmentation}

Object detection is a computer vision technology used to identify and locate specific targets within an image. In medical image segmentation, it is applied to automatically identify structures such as lesions, tumors, and organs, thereby assisting clinicians in diagnosis and treatment planning. Despite its widespread use in the literature, applications of object detection-assisted segmentation in LN segmentation remain relatively limited.

Existing studies, such as those by Bouget~\etal~\cite{semantic_bouget_2019} and Zhao~\etal~\cite{mri_zhao_2020}, have utilized Mask R-CNN~\cite{maskrcnn_he_2017} to simultaneously perform detection and segmentation. Mask R-CNN has proven effective in generating precise pixel-level segmentation masks. However, its performance is often constrained by the need for large-scale annotated datasets. Recent advances in detection-based segmentation have introduced models such as YOLOv8~\cite{yolov8_reis_2023}, Faster R-CNN~\cite{fasterrcnn_ren_2016}, and DETR~\cite{detr_carion_2020}, which have shown promising improvements in object detection and segmentation tasks in various domains.

For instance, YOLOv8, with its single-stage architecture, offers faster inference speeds and can potentially streamline real-time LN detection. Faster R-CNN continues to demonstrate strong performance in two-stage detection frameworks, while DETR's transformer-based approach allows for effective global context modeling, which may improve detection accuracy in complex anatomical structures. Integrating these modern techniques either as independent models or within hybrid architectures could address some of the limitations inherent in traditional methods and enhance overall performance in LN segmentation.

Future research should concentrate on comparing these approaches with established methods, investigating their adaptability to limited annotated data, as well as testing the performance of pre-trained models fine-tuned on small medical datasets and exploring how hybrid architectures can take advantage of detection and segmentation networks to improve clinical outcomes.

\subsection{Loss function of segmentation}

In addition to the development of more accurate and innovative LN segmentation deep learning models, previous studies have also addressed the construction of loss functions. In the context of LN segmentation, the construction of loss functions is important in optimizing the accuracy and performance during LN delineation. The incorporation of advanced loss functions, including focal loss, Dice loss, and cross-entropy loss, has the potential to markedly enhance segmentation outcomes. This is due to their effectiveness in handling pixel-level classification, optimizing the overlap between predicted and true masks, addressing class imbalance and consequently improving segmentation accuracy and model robustness.

In recent advancements, Xu~\etal~\cite{focal_xu_2020, atte_loss_xu_2020} introduced innovative approaches to enhance the segmentation of pathological LNs in PET/CT images through the development of specialized loss functions. Initially, they proposed a boundary-attention cross-entropy (BCE) loss function that focuses on increasing the weight of LN boundary voxels to address the challenge of accurately delineating LN boundaries in complex anatomical structures. Furthermore, they explored the integration of multiple loss functions, including BCE with generalized Dice loss, to effectively handle class imbalance and small-size issues associated with pathological LNs. This multifaceted approach was tested on architectures such as SegNet and DeepLabv3+, showing significant improvements in segmentation accuracy, as evidenced by high sensitivity and Dice scores, highlighting the potential of tailored loss functions in medical image segmentation tasks.

\section{Discussion}\label{sec:discussion}

\begin{figure*}[htb]
	\centering
	\includegraphics[width=0.9\textwidth]{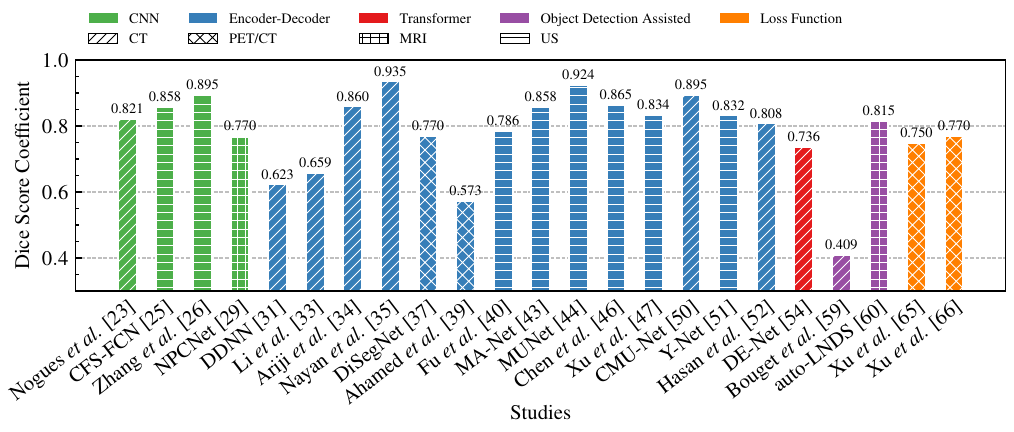}
	\caption{Overview of Dice scores of included studies.}
	\label{fig:dsc_overview}
\end{figure*}

From the above overview of the different studies, the key contributions of these different approaches and the corresponding experimental quantitative assessment results are summarized in Table~\ref{tab:summary_of_methods}. Due to the variability of evaluation metrics, we only present the common metric among all included studies, the Dice similarity coefficient (DSC, F1 score), as the primary metric for comparison. The overview of Dice scores of included studies is shown in Figure~\ref{fig:dsc_overview}. As observed in Figure~\ref{fig:dsc_overview}, certain methods~\cite{mediastinal_nayan_2022,munet_zhang_2020,decompose_zhang_2019} consistently outperform others, achieving higher Dice scores. These high-performing methods typically employ robust data augmentation techniques and utilize larger, more diverse datasets tend to demonstrate better generalization, resulting in higher Dice scores.

\renewcommand{\arraystretch}{1.1}

\begin{table*}[htbp]
	\caption{Summary of deep learning methods for LN segmentation.}
	\tiny
	\centering
	\setlength{\tabcolsep}{0.3em}
	\begin{tabular}{llL{0.24\linewidth}L{0.09\linewidth}lllllL{0.135\linewidth}l}
		\toprule
		Technique                                                        & Study                                        & Brief Overview                                                                                                                                                                                                                                                                         & Backbone                                                                                                                                                                       & Dataset                                                                                                  & Dataset Size                                                       & Modality & LN Site  & Image Size     & Preprocessing ($\mathcal{P}$) \& Augmentations ($\mathcal{A}$)                                                                                                                               & Performance                                                                           \\
		\midrule
		\multirow{4}{*}[-4em]{CNN}                                       & Nogues~\etal~\cite{holistically_nogues_2016} & Combined holistically-nested neural network (HNN) and boundary neural fields (BNF) together.                                                                                                                                                                                           & HNN~\cite{hnn_xie_2015} pre-trained on ImageNet~\cite{imagenet_deng_2009}                                                                                                      & TCIA~\cite{tcia_roth_2018}                                                                               & Total: 39,361                                                      & CT       & TA       & Uns.           & Uns.                                                                                                                                                                                         & \makecell[l]{DSC: 0.821\\IoU: 0.706\\mRVD: 0.137}                                     \\ \cline{2-11}
		                                                                 & CFS-FCN~\cite{coarse_zhang_2016}             & (1) Proposed coarse-to-fine stacked fully convolutional nets; (2) Developed boundary refinement method as post-processing.                                                                                                                                                             & FCN~\cite{fcn_long_2015}                                                                                                                                                       & Private                                                                                                  & Total: 80                                                          & US       & Uns.     & 388$\times$388 & Uns.                                                                                                                                                                                         & \makecell[l]{DSC: 0.858\\IoU: 0.860}                                                  \\ \cline{2-11}
		                                                                 & Zhang~\etal~\cite{decompose_zhang_2019}      & The original segmentation task was decomposed by classes or shapes of ROIs to multiple sub-tasks, then integrated for model training.                                                                                                                                                  & CUMedNet~\cite{cumednet_chen_2016}                                                                                                                                             & Private                                                                                                  & \makecell[l]{Train: 137\\Test: 100}                                & US       & Uns.     & 192$\times$192 & $\mathcal{A}$: random cropping, rotation and flipping.                                                                                                                                       & \makecell[l]{DSC: 0.895\\IoU: 0.810\\Prec: 0.901\\Rec: 0.889}                         \\ \cline{2-11}
		                                                                 & NPCNet~\cite{npcnet_li_2022}                 & Proposed position enhancement module (PEM), scale enhancement module (SEM), and boundary enhancement module (BEM) to tackle the variable location, variable size, and irregular boundary challenge.                                                                                    & ResNet-101~\cite{resnet_kaiming_2016}                                                                                                                                          & Private                                                                                                  & \makecell[l]{Train: 7,300\\Test: 1,824}                            & MRI      & H\&N     & 512$\times$512 & $\mathcal{P}$: resizing to 512$\times$512 pixels and min-max normalization.\newline $\mathcal{A}$: random rotation and flipping.                                                             & \makecell[l]{DSC: 0.770\\Prec: 0.800\\Rec: 0.780}                                     \\
		\midrule
		\multirow{12}{*}[-20em]{\makecell[l]{Encoder-\\Decoder}}         & DDNN~\cite{ddnn_men_2017}                    & Proposed U-shape model to segment nasopharynx gross tumor volume (GTVnx), the metastatic LN gross tumor volume (GTVnd) and the clinical target volume (CTV) simultaneously.                                                                                                            & \makecell[l]{VGG-16~\cite{vgg_simonyan_2014}\\DeConv~\cite{deconv_noh_2015}}                                                                                                   & Private                                                                                                  & \makecell[l]{Train: 184\\Test: 46}                                 & CT       & H\&N     & 417$\times$417 & $\mathcal{A}$: random cropping and flipping.                                                                                                                                                 & \makecell[l]{DSC: 0.623\\HD: 25.8~mm}                                                 \\ \cline{2-11}
		                                                                 & Li~\etal~\cite{tumor_li_2019}                & Modified U-Net to segment nasopharyngeal carcinoma (NPC) tumor targets at different stages.                                                                                                                                                                                            & U-Net~\cite{unet_ronneberger_2015}                                                                                                                                             & Private                                                                                                  & \makecell[l]{Train: 13,310\\Validation: 3,673\\Test: 3,693}        & CT       & H\&N     & 224$\times$224 & $\mathcal{P}$: cropping to 224$\times$224 pixels and min-max normalization.                                                                                                                  & \makecell[l]{DSC: 0.659\\HD: 32.10~mm}                                                \\ \cline{2-11}
		                                                                 & Ariji~\etal~\cite{seg_ariji_2022}            & Metastatic and non-metastatic cervical LNs from contrast-enhanced CT images are segmented by U-Net, and the performance was compared with radiologists.                                                                                                                                & U-Net~\cite{unet_ronneberger_2015}                                                                                                                                             & Private                                                                                                  & \makecell[l]{Train: 834\\Validation: 77\\Test: 72}                 & CT       & Neck     & 512$\times$512 & Uns.                                                                                                                                                                                         & \makecell[l]{DSC: 0.860\\Prec: 0.975\\Rec: 0.769}                                     \\ \cline{2-11}
		                                                                 & Nayan~\etal~\cite{mediastinal_nayan_2022}    & Modified UNet++ with bilinear interpolation (to avoid artifacts) and total generalized variation (TGV, to denoising).                                                                                                                                                                  & UNet++~\cite{unet++_zhou_2018}                                                                                                                                                 & \makecell[l]{TCIA~\cite{tcia_li_2020}\\5-Patients~\cite{5-patients_2022}\\ELCAP~\cite{lung_armato_2015}} & \makecell[l]{Train: 28,830\\Test: 25,500}                          & CT       & Lung     & 512$\times$512 & $\mathcal{A}$: random cropping, affine modification, flipping, noise and blur reduction, contrast and brightness controlling.                                                                & \makecell[l]{DSC: 0.935\\IoU: 0.919\\Acc: 0.948\\Prec: 0.931\\Rec: 0.941}             \\ \cline{2-11}
		                                                                 & DiSegNet~\cite{disegnet_xu_2021}             & (1) Proposed a new loss function called cosine-sine loss; (2) Combined multi-stage and multi-scale atrous spatial pyramid pooling sub-module (MS-ASPP).                                                                                                                                & SegNet~\cite{segnet_badrinarayanan_2017}                                                                                                                                       & Private                                                                                                  & \makecell[l]{Train: 3,710\\Test: 700}                              & PET/CT   & Thorax   & 256$\times$256 & $\mathcal{A}$: random translation in horizontal and vertical directions ($\pm$10 voxels).                                                                                                    & DSC: 0.770                                                                            \\ \cline{2-11}
		                                                                 & Ahamed~\etal~\cite{seg_ahamed_2023}          & Slicing 3D images into multiple 2D images, segmenting the primary tumors (GTVp) and metastatic LNs (GTVn) by 2D model separately and then re-stacking them as 3D images.                                                                                                               & U-Net~\cite{unet_ronneberger_2015} with ResNet-50~\cite{resnet_kaiming_2016} pre-trained on ImageNet~\cite{imagenet_deng_2009}                                                 & HECKTOR~\cite{hecktor_2022}                                                                              & \makecell[l]{Train: 524\\Test: 329}                                & PET/CT   & H\&N     & 128$\times$128 & $\mathcal{P}$: 5$\times$5$\times$5 median filtering, resizing to 128$\times$128 pixels.                                                                                                      & DSC: 0.573                                                                            \\ \cline{2-11}
		                                                                 & Fu~\etal~\cite{multi_modal_fu_2020}          & (1) Proposed a multi-modal fusion method for LN segmentation tasks; (2) Combined ultrasound and Doppler modalities to detect cervical LNs; (3) Applied the attention mechanism to the input feature map.                                                                               & U-Net~\cite{unet_ronneberger_2015}                                                                                                                                             & Private                                                                                                  & \makecell[l]{Train: 634\\Validation: 211\\Test: 209}               & US       & H\&N     & 256$\times$256 & $\mathcal{P}$: resizing to 256$\times$256$\times$3, 5$\times$5 median filtering and modalities registration.                                                                                 & DSC: 0.786                                                                            \\ \cline{2-11}
		                                                                 & MA-Net~\cite{manet_zhang_2020}               & Developed an encoder-decoder-like multiple-channel and atrous convolution network, including pyramid pooling and residual connections.                                                                                                                                                 & \makecell[l]{U-Net~\cite{unet_ronneberger_2015}\\Dilated CNN~\cite{dilated_yu_2015}}                                                                                           & Private                                                                                                  & \makecell[l]{Train: 160\\Test: 50}                                 & US       & Uns.     & 320$\times$256 & $\mathcal{P}$: cropping to 512$\times$512 pixels.\newline $\mathcal{A}$: horizontal and vertical flipping, random scaling ($\pm$10\%), random rotation (0 to 10°) and flipping.              & \makecell[l]{DSC: 0.858\\Prec: 0.854\\Rec: 0.885\\HD: 19.245~mm\\ASD: 4.312~mm}       \\ \cline{2-11}
		                                                                 & MUNet~\cite{munet_zhang_2020}                & Proposed a fully convolutional network with a replaceable backbone that accepts input images of arbitrary size.                                                                                                                                                                        & \makecell[l]{U-Net~\cite{unet_ronneberger_2015}\\ResNet-50/152~\cite{resnet_kaiming_2016}\\Inception v3/v4~\cite{inceptionv3_szegedy_2016}}                                    & Private                                                                                                  & \makecell[l]{Train: 4,000\\Validation: 1,000\\Test: 1,000}         & US       & Neck     & Multiple       & $\mathcal{P}$: resizing to the combination of \{256,384,512,640\}.\newline $\mathcal{A}$: random rotation ($\pm$5°), flipping and adding Gaussian white noise (variances of 0.0001 to 0.01). & \makecell[l]{DSC: 0.924\\Acc: 0.932}                                                  \\ \cline{2-11}
		                                                                 & Chen~\etal~\cite{diffusion_chen_2021}        & (1) Adopted Gabor-based anisotropic diffusion (GAD) to reduce speckle noise in ultrasound images; (2) Filled the hila portion in segmentation results using morphological operations.                                                                                                  & U-Net~\cite{unet_ronneberger_2015}                                                                                                                                             & Private                                                                                                  & \makecell[l]{Train: 390\\Validation: 51\\Test: 90}                 & US       & Uns.     & 240$\times$240 & $\mathcal{A}$: random flipping, shifting, rotation, shearing, brightness and contrast and elastic transformation.                                                                            & \makecell[l]{DSC: 0.865\\IoU: 0.763\\Acc: 0.934\\Sen: 0.939\\Spc: 0.937}              \\ \cline{2-11}
		                                                                 & Xu~\etal~\cite{difficulty_xu_2022}           & (1) A difficulty-aware module is proposed to distinguish the difficulty grade of LN images and apply corresponding segmentation branches according to the difficulties; (2) A spatial attention module is adopted to the complex segmentation branch.                                  & \makecell[l]{ResNet~\cite{resnet_kaiming_2016}\\ASPP~\cite{aspp_chen_2017}}                                                                                                    & Private                                                                                                  & \makecell[l]{Train: 1,200\\Test: 66}                               & US       & Armpit   & Uns.           & $\mathcal{A}$: random rotation and translation.                                                                                                                                              & \makecell[l]{DSC: 0.834\\IoU: 0.744\\VOE: 0.120}                                      \\ \cline{2-11}
		                                                                 & CMU-Net~\cite{efficient_wen_2024}            & (1) Proposed cascaded multi-heads UNet (CMU-net); (2) The classification results were multiplied with the corresponding segmentation network as a post-processing method.                                                                                                              & UNet++~\cite{unet++_zhou_2018}                                                                                                                                                 & Private                                                                                                  & \makecell[l]{Train: 120 cases\\Test: 40 cases}                     & CT       & Pelvic   & 512$\times$512 & Uns.                                                                                                                                                                                         & \makecell[l]{DSC\(_{avg}\): 0.895\\ASD\(_{avg}\): 0.647~mm\\HD95\(_{avg}\): 2.811~mm} \\ \cline{2-11}
		                                                                 & Y-Net~\cite{deep_zhao_2024}                  & (1) Validated the effectiveness of Y-Net model in lymph node segmentation and classification; (2) The four-level pyramid pooling module and pyramid spatial pooling blocks were proposed to achieve segmentation and classification simultaneously.                                    & U-Net~\cite{unet_ronneberger_2015}                                                                                                                                             & Private                                                                                                  & \makecell[l]{Train \& Test: 2,512\\Validation: 547}                & US       & Cervical & Uns.           & Uns.                                                                                                                                                                                         & DSC: 0.832                                                                            \\ \cline{2-11}
		                                                                 & Hasan~\etal~\cite{automated_hasan_2024}      & (1) Proposed an attention block for traditional U-Net to reduce loss of crucial information during down sampling in decoder; (2) Proposed S-Net for small lymph node segmentation; (3) Combined the Tversky loss, BCE, and IoU loss to learn the characteristics of small lymph nodes. & \makecell[l]{U-Net~\cite{unet_ronneberger_2015}\\S-Net~\cite{focusnet_gao_2019}}                                                                                               & Private                                                                                                  & \makecell[l]{Train: 18,054\\Validation: 4,463\\Test: 2,602}        & CT       & Cervical & Uns.           & $\mathcal{A}$: random rotation ($\pm$10°), random vertical flip, random brightness-contrast change and random gamma transformation.                                                          & DSC: 0.808                                                                            \\
		\midrule
		Transformer                                                      & DE-Net~\cite{denet_shi_2022}                 & Proposed a dual-encoder U-shape model named DE-Net (composed of parallel CNN and Swin Transformer).                                                                                                                                                                                    & \makecell[l]{ResNet~\cite{resnet_kaiming_2016}\\Swin Transformer~\cite{swint_liu_2021}}                                                                                        & Private                                                                                                  & \makecell[l]{Train: 30 scans\\Test: 10 scans}                      & CT       & All      & 512$\times$512 & $\mathcal{A}$: random erasure, scaling, distortion, rotation, vertical flip and noise.                                                                                                       & \makecell[l]{DSC: 0.736\\HD: 21.500~mm}                                               \\
		\midrule
		\multirow{3}{*}[-1em]{\makecell[l]{Object\\Detection\\Assisted}} & Bouget~\etal~\cite{semantic_bouget_2019}     & (1) Proposed a three-step 2D pipeline to perform both semantic segmentation and instance detection; (2) Used Mask R-CNN to detect various target structures in different sizes.                                                                                                        & \makecell[l]{U-Net~\cite{unet_ronneberger_2015}\\Mask R-CNN~\cite{maskrcnn_he_2017}}                                                                                           & 17-Patients~\cite{airway_reynisson_2015}                                                                 & Total: 17 scans                                                    & CT       & Lung     & 256$\times$256 & $\mathcal{A}$: random rotation ($\pm$20°), flipping,  affine transformation, intensity range clipping and rescaling.                                                                         & DSC: 0.409                                                                            \\ \cline{2-11}
		                                                                 & auto-LNDS~\cite{mri_zhao_2020}               & (1) Combined T2WI and DWI together to generate three-channel images; (2) Used Mask R-CNN to detect and segment LNs simultaneously.                                                                                                                                                     & \makecell[l]{ResNet-101~\cite{resnet_kaiming_2016}\\Mask R-CNN~\cite{maskrcnn_he_2017}}                                                                                        & Private                                                                                                  & \makecell[l]{Train: 5,694\\Test$_{in}$: 1,192\\Test$_{ex}$: 2,572} & MRI      & Pelvic   & 256$\times$256 & $\mathcal{P}$: cropping to 256$\times$256 pixels.\newline $\mathcal{A}$: random cropping, affine transformation, flipping, adding noise, blurring, contrast and brightness enhancement.      & \makecell[l]{DSC\(_{in}\): 0.820\\DSC\(_{ex}\): 0.810\\DSC\(_{avg}\): 0.815}          \\
		\midrule
		\multirow{2}{*}[-1em]{\makecell[l]{Loss\\Function}}              & Xu~\etal~\cite{focal_xu_2020}                & Proposed the focal loss from object detection task to the segmentation task.                                                                                                                                                                                                           & SegNet \cite{segnet_badrinarayanan_2017}, DeepLab v3+ \cite{deeplabv3+_chen_2018}, ResNet-18 \cite{resnet_kaiming_2016}, all pre-trained on ImageNet \cite{imagenet_deng_2009} & Private                                                                                                  & Total: 63 scans                                                    & PET/CT   & Thorax   & Uns.           & $\mathcal{A}$: Random translation in horizontal and vertical directions (0 to 10 pixels).                                                                                                    & \makecell[l]{DSC: 0.750\\Sen: 0.870}                                                  \\ \cline{2-11}
		                                                                 & Xu~\etal~\cite{atte_loss_xu_2020}            & Combined the voxel-level loss function like boundary-attention cross-entropy loss into area-level loss function (generalized dice loss).                                                                                                                                               & VGG-16 \cite{vgg_simonyan_2014}, SegNet \cite{segnet_badrinarayanan_2017}, DeepLab v3+ \cite{deeplabv3+_chen_2018}, all pre-trained on ImageNet \cite{imagenet_deng_2009}      & Private                                                                                                  & Total: 63 scans                                                    & PET/CT   & Thorax   & Uns.           & $\mathcal{A}$: Random translation in horizontal and vertical directions (0 to 10 pixels).                                                                                                    & \makecell[l]{DSC: 0.770\\Sen: 0.880}                                                  \\
		\bottomrule
	\end{tabular}
	\label{tab:summary_of_methods}

	\smallskip

	\raggedright
	\begin{spacing}{0.91}
		\scriptsize
		\textbf{Backbone} CNN: convolutional neural network, FCN: fully convolutional network, ResNet: residual network, VGG: visual geometry group, ASPP: atrous spatial pyramid pooling, DeConv: deconvolutional network. \textbf{Modality} CT: computed tomography, MRI: magnetic resonance imaging, PET: positron emission tomography, US: ultrasound; \textbf{LN Site} H\&N: head and neck, TA: thoracoabdominal, Uns: unspecified; \textbf{Performance} DSC: dice similarity coefficient, IoU: intersection over union, Prec: precision, Rec: recall, Acc: accuracy, Sen: sensitivity, Spc: specificity, HD: Hausdorff distance, ASD: average surface distance, \(in\): internal dataset, \(ex\): external dataset, and \(avg\): on average for multi-center study metrics.
	\end{spacing}
\end{table*}

\renewcommand{\arraystretch}{1.0}

\subsection{Performance between techniques}

\begin{figure}[htbp]
	\centering
	\includegraphics[width=0.4\textwidth]{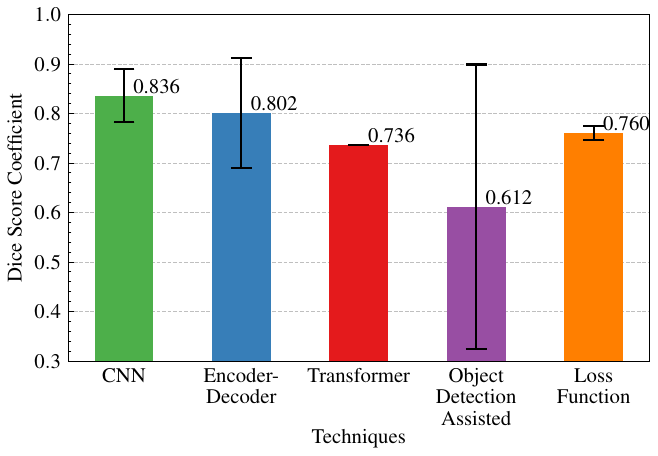}
	\caption{Mean and standard deviation of Dice scores for different techniques.}
	\label{fig:performance_between_techniques}
\end{figure}

As Table~\ref{tab:summary_of_methods} and Figure~\ref{fig:dsc_overview} indicate, the encoder-decoder architecture is currently widely applied to LN segmentation. It has been shown to be more scalable and efficient because of its straightforward design and comparative lower computational needs, making it particularly well-suited for real-time clinical applications. Furthermore, FCN enables the model to process images with arbitrary resolution, thereby enhancing its adaptability to medical data. In contrast, Transformer-based methods have seen limited adoption due to their substantial computational demands. The mean and standard deviation of Dice scores of different techniques are categorized and shown in Figure~\ref{fig:performance_between_techniques}. The results indicate that the performance of different techniques varies significantly, with the best-performing methods achieving Dice scores that are more than twice as high as the worst-performing methods.

Notably, Nayan~\etal~\cite{mediastinal_nayan_2022} achieved the best segmentation results on three public datasets for LN segmentation by using a modified UNet++, scoring 0.935 for DSC and 0.919 for IoU. The dataset used in Nayan~\etal~\cite{mediastinal_nayan_2022} is the largest dataset among the studies included in this review, with 54,330 images among three public datasets, and with a relatively high resolution of 512$\times$512. The extensive size of dataset likely contributed to the robustness and generalizability of the model. Additionally, the augmentation techniques employed, such as random cropping, flipping and contrast and brightness controlling, enhanced the diversity of the training data, further improving the capacity of model to generalize to unseen data. These factors collectively contributed to the superior performance of the model.

In contrast, the method proposed by Bouget~\etal~\cite{semantic_bouget_2019} achieved the lowest Dice score of 0.409 among the studies included. This result may be attributed to several factors. Firstly, the study used an original U-Net and Mask R-CNN without proper modification, and the U-Net was trained from scratch while the Mask R-CNN was pre-trained on the ImageNet dataset. This may lack the generalization ability of the model. Secondly, the objective of this study is not only focused on LN segmentation but also includes the segmentation of 14 other tissues and organs. This broader scope may have resulted in a relatively limited emphasis on the specific task of LN segmentation. Lastly, the input images generated from CT slices were resized to 256$\times$256, which may have led to the detailed information loss of LN structures, thereby affecting the LN segmentation performance.

As shown in Figure~\ref{fig:performance_between_techniques}, the highest Dice score was achieved by CNN-based methods (0.836), followed by encoder-decoder-based methods (0.802), and Transformer-based methods (0.736). This indicates that CNN-based methods are currently the most accurate for LN segmentation tasks. Meanwhile, although the Transformer architecture is widely used in the domain of natural image processing, it has not been extensively adopted for LN segmentation tasks. Preliminary research on Transformer-based DE-Net~\cite{denet_shi_2022}, utilizing Swin Transformer, achieved a Dice score of 0.736 on a private dataset. This indicates there is still significant room for improvement compared to the CNN and encoder-decoder architecture. Furthermore, it underscores the substantial potential for further development of Transformer technology in this task.

The results also show that the object detection-assisted segmentation methods have the lowest mean Dice score (0.612), which is significantly lower than the other methods. This may be because the object detection-assisted segmentation methods are not specifically designed for LN segmentation tasks, and the model may not be well-suited for the segmentation of small and irregularly shaped LN structures. Additionally, the object detection models that are used in these methods are pre-trained on the object detection natural image datasets and did not fine-tune using LN data, this may lead to less effectiveness in capturing the detailed information of complex structures, which could result in performance degradation. Furthermore, the limited number of studies employing object detection-assisted segmentation methods constrains the representativeness of the mean and standard deviation of Dice scores.

The studies that focused on improving the loss function of LN segmentation achieved a mean Dice score (0.760). These studies utilized older backbones such as SegNet and DeepLabv3+, in contrast to the popular U-Net, UNet++, and other encoder-decoder architectures. Additionally, the dataset used in these studies was relatively small, with an average of 63 images, which limited the model generalizability. Consequently, they exhibited lower performance compared to CNN and encoder-decoder-based methods.

\subsection{Performance between modalities}

\begin{figure}[htbp]
	\centering
	\includegraphics[width=0.4\textwidth]{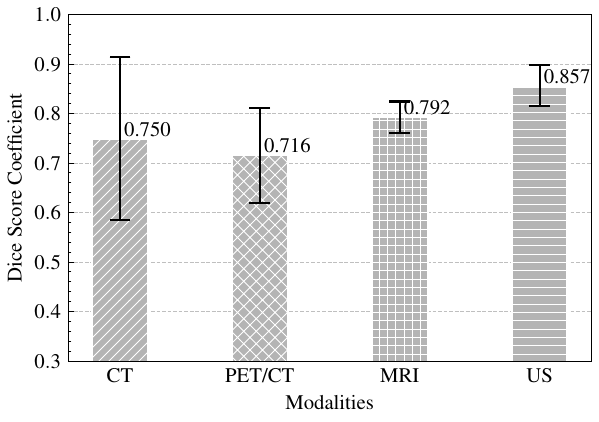}
	\caption{Mean and standard deviation of Dice scores for different modalities.}
	\label{fig:performance_between_modalities}
\end{figure}

Figure~\ref{fig:performance_between_modalities} shows the mean and standard deviation of Dice scores between different modalities. The results vary significantly, the performance of CT and PET/CT are relatively close, scoring a mean 0.750 and 0.716 Dice score, respectively. In contrast, the group of ultrasound modality achieved the highest mean Dice score of 0.857.

This phenomenon may be attributed to the image characteristics of different modalities and data processing methods. Due to the principles of CT and PET imaging, a LN occupies only a small area in CT and PET images, making it difficult to be delineated, detected, and segmented. Meanwhile, the resolution and contrast of raw images generated by CT and PET are various, and preprocessing techniques including resizing, histogram equalization, and others are essential to ensure the consistency of the input data. This may result in the loss of detailed information on LN structures, thereby affecting the LN segmentation performance.

Although the signal-to-noise ratio (SNR) of ultrasound images is lower compared to CT and PET images, ultrasound imaging is highly effective in displaying LN during real-time examinations. The ability to zooming in, zooming out, and manipulating the probe to observe the LN from different angles, providing a comprehensive view of the entire LN structure. Additionally, ultrasound images captured by radiologists often have a higher proportion of LN compared to CT and PET images, further aiding in segmentation. Furthermore, high-frequency ultrasound is commonly used in LN imaging. The relatively high image resolution of high-frequency ultrasound allows preprocessing operations and enhances the potential for high-accuracy segmentation.

The performance of the MRI modality achieved a balance between CT/PET and ultrasound, with a mean Dice score of 0.792. MRI is widely used in clinical practice due to its excellent soft tissue contrast and multiplanar imaging capabilities. The high resolution and contrast of MRI images enable the capture of detailed information on LN structures. However, the MRI modality is least utilized in the studies included in this study, possibly due to the high cost of MRI imaging and the complexity of MRI image processing. Further exploration is needed to fully understand the potential performance of MRI in LN segmentation tasks.

\subsection{Best method for different modalities}

According to the previous literature, the best method for different modalities in LN segmentation are shown in Table~\ref{tab:best_methods_for_different_modalities}. The results indicate that the best methods for CT, PET/CT, and ultrasound are encoder-decoder-based methods, while the best method for the MRI modality is an object detection-assisted method.

\begin{table*}[htbp]
	\caption{Best methods for different modalities in LN segmentation. The subscript \(in\), and \(ex\) indicate the internal dataset and external dataset.}
	\footnotesize
	\centering
	\begin{tabular}{llllllll}
		\toprule
		Modality & Study                                     & Technique                                 & Backbone                                                                                                                                    & Dataset                                                                                                  & Dataset Size                                                           & LN Site & Performance                                                                  \\
		\midrule
		CT       & Nayan~\etal~\cite{mediastinal_nayan_2022} & Encoder-Decoder                           & UNet++~\cite{unet++_zhou_2018}                                                                                                              & \makecell[l]{TCIA~\cite{tcia_li_2020}\\5-Patients~\cite{5-patients_2022}\\ELCAP~\cite{lung_armato_2015}} & \makecell[l]{Train: 28,830 \\ Test: 25,500}                            & Lung    & \makecell[l]{DSC: 0.935\\IoU: 0.919\\Acc: 0.948\\Prec: 0.931\\Rec: 0.941}    \\
		\midrule
		PET/CT   & DiSegNet~\cite{disegnet_xu_2021}          & Encoder-Decoder                           & SegNet~\cite{segnet_badrinarayanan_2017}                                                                                                    & Private                                                                                                  & \makecell[l]{Train: 3,710 \\ Test: 700}                                & Thorax  & DSC: 0.770                                                                   \\
		\midrule
		MRI      & auto-LNDS~\cite{mri_zhao_2020}            & \makecell[l]{Object Detection\\ Assisted} & \makecell[l]{ResNet 101~\cite{resnet_kaiming_2016}\\Mask R-CNN~\cite{maskrcnn_he_2017}}                                                     & Private                                                                                                  & \makecell[l]{Train: 5,694 \\ Test$_{in}$: 1,192 \\ Test$_{ex}$: 2,572} & Pelvic  & \makecell[l]{DSC\(_{in}\): 0.820\\DSC\(_{ex}\): 0.810\\DSC\(_{avg}\): 0.815} \\
		\midrule
		US       & MUNet~\cite{munet_zhang_2020}             & Encoder-Decoder                           & \makecell[l]{U-Net~\cite{unet_ronneberger_2015}\\ResNet 50/152~\cite{resnet_kaiming_2016}\\Inception v3/v4~\cite{inceptionv3_szegedy_2016}} & Private                                                                                                  & \makecell[l]{Train: 4,000 \\ Validation: 1,000 \\ Test: 1,000}         & Neck    & \makecell[l]{DSC: 0.924\\Acc: 0.932}                                         \\
		\bottomrule
	\end{tabular}
	\label{tab:best_methods_for_different_modalities}

	\smallskip

	\raggedright
	\begin{spacing}{0.91}
		\scriptsize
		\textbf{Modality} CT: computed tomography, MRI: magnetic resonance imaging, PET: positron emission tomography, US: ultrasound; \textbf{Backbone} ResNet: residual network; \textbf{Performance} DSC: dice similarity coefficient, IoU: intersection over union, Prec: precision, Rec: recall, Acc: accuracy, \(in\): internal dataset, \(ex\): external dataset, and \(avg\): on average for multi-center study metrics.
	\end{spacing}
\end{table*}

Interestingly, the majority of studies included in this study have utilized U-Net, variants of U-Net such as UNet++, and ResNet, as the backbone architectures. These backbones have established themselves as highly effective in medical image segmentation tasks. The fact that the best methods for different modalities also rely on these backbones further highlights their robustness and generalizability in LN segmentation tasks.

Analysis of Table~\ref{tab:best_methods_for_different_modalities} and the model architectures of these methods (\cite{mediastinal_nayan_2022,disegnet_xu_2021,mri_zhao_2020,munet_zhang_2020}) reveals three common characteristics that significantly contribute to their superior performance across different modalities: (1) implementation of encoder-decoder architectures with skip connections, enabling effective capture of both local details and global contextual features; (2) training on relatively large and diverse datasets, enhancing model generalization; and (3) extensive use of data augmentation techniques to artificially expand the training data variability. These complementary strategies collectively facilitate robust feature extraction and accurate segmentation performance, regardless of the underlying imaging modality.

\subsection{Overall observations}

\subsubsection{Advantages}

The application of deep learning techniques in LN segmentation tasks offers three principal advantages. Firstly, it has the potential to considerably reduce the time and effort required by radiologists for diagnosis. It facilitates the optimization of the clinical workflow by automating the identification of metastatic nodes, thereby enhances the efficiency of LN image analysis. Secondly, deep learning methodologies can assist in standardizing the process, and ensure consistency in the evaluation and reduce the variability in accuracy that may arise from the differences in experience and expertise amongst radiologists. The automation of the segmentation process allows models to achieve more reliable and reproducible results, and provides real-time support for diagnostic and therapeutic decisions. Third, the continuous learning and transfer learning capabilities enable the integration of new data and knowledge, facilitates the development of more accurate and robust models~\cite{continuous_wang_2024}. By leveraging the vast amount of data and information available, deep learning models can enhance the performance and generalizability of LN segmentation tasks.

\subsubsection{Limitations}

At present, the most frequently utilized deep learning architectures for LN segmentation are encoder-decoder-based methods, particularly U-Net and its variants. Although these methods have demonstrated high performance in various medical image segmentation tasks, they may not be the most suitable methods due to the bias attributed to the different private datasets used in the studies. The insufficiency of publicly accessible datasets and the scarcity of diversity in the data utilized for training may have resulted in inequitable comparisons between disparate methodologies.

Additionally, the limited diversity of datasets in terms of patient demographics, imaging protocols, and anatomical variations poses challenges for models to generalize well across different clinical settings. The quality of annotations exhibits variability across studies, with some relying on less detailed or inconsistent labeling, which can adversely affect the training and evaluation of segmentation models. These issues underscore the need for more standardized and diverse datasets, as well as higher annotation quality, to advance the field and enable fairer comparisons of different methodologies.

Moreover, the performance of Transformer-based techniques and object detection-assisted segmentation methods in LN segmentation tasks is comparatively inferior to that of CNN-based methods. This may hamper the widespread application of these techniques in LN segmentation tasks and hinder the full realization of their potential.

\subsubsection{Challenges}

\textbf{Data scarcity and annotation complexity.} It is worth noting that out of the 23 studies included in this study, only four studies~\cite{holistically_nogues_2016,mediastinal_nayan_2022,seg_ahamed_2023,semantic_bouget_2019} utilized public datasets~\cite{tcia_roth_2018,tcia_li_2020,5-patients_2022,lung_armato_2015,hecktor_2022,airway_reynisson_2015}, while the remaining studies relied on private datasets. The number of images used in each study also varied. The use of different datasets makes it challenging to directly compare the experimental results of various studies and ascertain the superiority of methods, and this situation hampers the innovative development of LN segmentation tasks.

In recent years, with the availability of numerous publicly accessible datasets such as ImageNet~\cite{imagenet_deng_2009}, COCO~\cite{coco_lin_2014}, and Cityscapes~\cite{cityscapes_cordts_2016}, natural vision tasks have experienced significant advancements. However, medical imaging tasks, including those related to LN image processing, have faced persistent challenges in data availability. Difficulties in data collection, high acquisition costs, and the labor-intensive nature of data annotation have hindered the formation of large datasets in this domain.

Acquiring and annotating medical imaging data is complex. Devices from different manufacturers or time periods vary in resolution, contrast, noise, and artifacts due to technological advancements and different settings. These variations affect image quality and the amount of usable information, making it difficult to build deep learning datasets. Inconsistent data from different devices complicates preprocessing and standardization, reducing model generalization and accuracy. In addition, device incompatibilities limit data integration from multiple sources, resulting in small and heterogeneous datasets that weaken the performance and reliability of deep learning models in clinical settings. Therefore, effective calibration and standardization measures should be implemented to ensure data consistency and quality during data acquisition, thereby improving the diagnostic accuracy and clinical utility of deep learning models.

Unlike the annotating process of natural images, the delineation process of LN structures highly depends on the expertise of radiologists. It takes several minutes to annotate a single LN image for professional radiologists, making the collection and annotation of a large dataset of LN images time-consuming and costly. Due to these challenges, researchers often rely on limited self-collected data, which is constrained by ethical and copyright restrictions imposed on hospital data. Consequently, the scarcity of data hampers research efficiency and imposes limitations on the performance of models in LN segmentation tasks.

Moreover, LNs exhibit considerable variability in morphology and location across different patients and imaging modalities. This diversity increases the difficulty of automatic segmentation, requiring models to have strong generalization capabilities to perform reliably in various clinical environments. Additionally, consistency in annotation is also an issue, as different annotators might have varying delineations for the same LN, introducing noise into the dataset.

\textbf{Insufficient application of new techniques.} Another challenge is the insufficient application of new techniques such as object detection-assisted methods, Transformer models, and large-scale pre-trained models. The object detection-assisted methods included in this study only adopted the Mask R-CNN model which is pre-trained on the ImageNet dataset. The lack of fine-tuning on medical images may have limited the performance of these methods. In addition, while transformers and large-scale pre-trained models are widely used in natural language and image processing tasks, their application in LN segmentation tasks is still scarce. Integrating these models has the potential to improve the performance and generalization of models in LN segmentation tasks.

\textbf{Clinical integration challenges.} The integration of deep learning-based segmentation models into clinical workflows presents several challenges. A primary concern is the seamless integration with existing clinical systems, such as Picture Archiving and Communication Systems (PACS). The successful incorporation of these tools requires interoperability that ensures the outputs are directly accessible and interpretable within the established clinical infrastructure without disrupting the standard radiological workflow. In addition to the integration of technology, the regulatory environment presents challenges. Regulatory agencies such as the U.S. Food and Drug Administration (FDA) and the European Conformity Assessment (CE) bodies require rigorous evidence of safety, efficacy, and reproducibility prior to approval.

Moreover, the implementation of these models in actual settings is hindered by the inherent variations among healthcare institutions, such as imaging protocols, hardware configurations, and data management practices. Consequently, deep learning models must possess a high degree of adaptability and be thoroughly validated in diverse clinical environments. The effective integration of such protocols necessitates robust multidisciplinary collaboration. The convergence of expertise from clinicians, biomedical engineers, information technology specialists, and regulatory experts is imperative for the development of standardized protocols that adhere to technical and regulatory requirements, and ensure the seamless integration of the tools into routine clinical practice.

\textbf{Uncertainty quantification.} An important aspect not previously addressed is the quantification of uncertainty in deep learning-based segmentation. In clinical applications, it is crucial that segmentation models provide predictions and confidence estimates. Techniques such as Monte Carlo dropout, which involves performing multiple stochastic forward passes during inference, and Bayesian approximation approaches that model parameter uncertainty, can be employed to quantify uncertainty. In addition, depth ensembles quantify uncertainty through the variance of the prediction results, offering another avenue to assess prediction variability.In the context of LN segmentation, providing uncertainty estimates may enable clinicians to identify regions with lower confidence, prompting further review or alternative diagnostic measures.

\subsection{Future perspectives}

In response to the current challenges faced by LN segmentation tasks, we propose several potential research directions for future development.

\textbf{Workaround for insufficient data.} To address the issue of limited data availability, researchers can adopt several strategies. Firstly, while conducting experiments on private datasets, it is also important to evaluate model performance on publicly available datasets, such as the Sa-med2d-20m dataset~\cite{sam_dataset_ye_2023}. This allows for broader participation and facilitates the improvement of methods. Secondly, researchers should actively explore new techniques that are less reliant on large volumes of data. Approaches such as transfer learning~\cite{mri_zhao_2020}, semi-supervised learning~\cite{bcp_bai_2023}, weakly supervised learning~\cite{ccam_chen_2022}, contrastive learning~\cite{byol_grill_2020} and unsupervised learning~\cite{unsupervised_chen_2020} can enable the construction of models with minimal data. These methods leverage existing prior knowledge and can effectively enhance the performance of models in LN segmentation tasks.

The application of diverse data transformations, including rotation, flipping, scaling, and contrast adjustment, to existing images facilitates the augmentation of training data, thereby enhancing the diversity of the training data set and simulating the variations in images that arise from different devices and imaging conditions. Furthermore, data augmentation can emulate the anatomical and pathological variations observed in real-world scenarios, thereby increasing the robustness of deep learning models when processing images from diverse sources and of varying quality. Consequently, data augmentation techniques are pivotal in improving the accuracy and reliability of models and in promoting the effectiveness of deep learning algorithms in clinical applications.

\textbf{Model hybridization.} The combination of multiple models has been proven an effective approach to image segmentation, such as CNNs and transformers. As demonstrated by the DETR~\cite{detr_carion_2020}, this integration capitalizes on the respective strengths of these architectures. CNNs are capable at capturing local features, whereas transformers excel at learning global representations. The synergy between these models enhances segmentation accuracy, improves generalization, optimizes computational efficiency by reducing the workload on transformers, and increases model robustness against noise and variability. Despite the greater computational demands of Transformer models, strategies such as quantization, parameter pruning, and lightweight Transformer (\eg, TinyViT~\cite{tiny_vit_wu_2022}) can be employed to alleviate the burden and enhance the feasibility of model hybridization. This integration of diverse models may has the potential to the development of more precise and reliable LN segmentation methods.

\textbf{Multimodal fusion techniques.} To advance LN segmentation, it is essential to expand the application of multimodal fusion technologies, such as combining grayscale ultrasound images with color Doppler images~\cite{multi_modal_fu_2020}, integrating CT and PET images~\cite{weighted_mahdi_2024}, and merging multiple imaging modalities like CT and MRI with patient clinical information. This integration provides richer blood flow data, detailed tissue structures, and comprehensive patient backgrounds, enabling deep learning models to achieve more accurate and robust segmentation. By leveraging multiple data sources, models can better generalize across diverse clinical scenarios, reducing the likelihood of missed or incorrect detections. Additionally, incorporating clinical information enhances the clinical relevance and practicality of segmentation results, supports personalized treatment plans, and improves diagnostic confidence.

\textbf{Integration with large-scale pre-trained models.} Exploring the integration with large-scale pre-trained models, such as GPT~\cite{gpt_brown_2020} and the Segment Anything Model (SAM)~\cite{sam_kirillov_2023,sam_med2d_cheng_2023}, which are trained on vast datasets, has become increasingly relevant in the medical diagnosis field~\cite{collaborative_wu_2024}. This is important for researchers lacking sufficient data or computational resources. Utilizing large models for fine-tuning in specific domains has emerged as a new trend and direction. By leveraging the knowledge learned from these pre-trained models, researchers can potentially improve the performance and generalizability of LN segmentation models, even with limited data. Furthermore, the integration of pre-trained models can also facilitate the transfer of learned features and representations, enabling the development of more robust and accurate LN segmentation methods.

\subsection{Limitations}

This systematic review is subject to four limitations.

\begin{enumerate}
	\item Commonly used evaluation metrics across the included studies are limited and compromise the comparability of their results; while DSC is the most commonly reported metric, it primarily reflects the overlap between predicted and ground truth regions but provides limited insight into boundary accuracy. This reliance on DSC may overlook important aspects of segmentation performance, particularly in clinical contexts where accurate boundary delineation is critical. Metrics such as HD (sensitive to outlier boundary points), ASD (captures the overall contour alignment), and VOE (quantifies volumetric discrepancies), offer a more comprehensive understanding of segmentation performance. However, these metrics were not consistently reported across the included studies, limiting their use in our analysis. Future studies should report a wider range of performance metrics to enable a more holistic comparison.
	\item The exclusion criteria, which omitted studies lacking detailed information on deep learning techniques and quantitative results, may have limited the comprehensiveness of the review.
	\item The utilization of datasets with considerable variations in quantity and quality may introduce biases in method comparisons.
	\item Few studies included in Transformer and object detection-assisted group, may result in an incomplete representation of the field.
\end{enumerate}

These limitations may affect the generalizability of the conclusions, restricting their applicability to studies that meet these specific criteria and reducing the broader transferability of the review findings.

\section{Conclusion}\label{sec:conclusion}

This study offers a comprehensive overview of deep learning techniques applied to LN segmentation tasks. It analyzes the performance of various techniques in multiple LN imaging modalities, identifying optimal methods for each modality. Analyses from different perspectives show that deep learning methods provide comparable results to manual segmentation, especially in large datasets. Deep learning methods also enable efficient full automation process, helping to streamline the clinical diagnostic workflows. The challenges and limitations currently hindering progress in this research area are thoroughly discussed. Additionally, potential directions for future research are proposed. By summarizing the current state of research, this study provides valuable insights for LN segmentation researchers and contributes to advancing medical image processing.

\section*{Conflict of interest statement}

The authors declare no conflict of interest.

\section*{Acknowledgment}

This work was supported by a General Research Fund of the Research Grant Council of Hong Kong (Reference no. 15102222).

\bibliographystyle{IEEEtran}
\bibliography{ref}

\begin{thebibliography}{10}
\providecommand{\url}[1]{#1}
\csname url@samestyle\endcsname
\providecommand{\newblock}{\relax}
\providecommand{\bibinfo}[2]{#2}
\providecommand{\BIBentrySTDinterwordspacing}{\spaceskip=0pt\relax}
\providecommand{\BIBentryALTinterwordstretchfactor}{4}
\providecommand{\BIBentryALTinterwordspacing}{\spaceskip=\fontdimen2\font plus
\BIBentryALTinterwordstretchfactor\fontdimen3\font minus \fontdimen4\font\relax}
\providecommand{\BIBforeignlanguage}[2]{{%
\expandafter\ifx\csname l@#1\endcsname\relax
\typeout{** WARNING: IEEEtran.bst: No hyphenation pattern has been}%
\typeout{** loaded for the language `#1'. Using the pattern for}%
\typeout{** the default language instead.}%
\else
\language=\csname l@#1\endcsname
\fi
#2}}
\providecommand{\BIBdecl}{\relax}
\BIBdecl

\bibitem{lymphadenopathy_bazemore_2002}
A.~W. Bazemore and D.~R. Smucker, ``Lymphadenopathy and malignancy,'' \emph{Am Fam Physician}, vol.~66, no.~11, pp. 2103--10, 2002.

\bibitem{ultrasound_ahuja_2008}
\BIBentryALTinterwordspacing
A.~T. Ahuja, M.~Ying, S.~Y. Ho, G.~Antonio, Y.~P. Lee, A.~D. King, and K.~T. Wong, ``Ultrasound of malignant cervical lymph nodes,'' \emph{Cancer Imaging}, vol.~8, no.~1, pp. 48--56, 2008. [Online]. Available: \url{https://www.ncbi.nlm.nih.gov/pubmed/18390388}
\BIBentrySTDinterwordspacing

\bibitem{sonographic_ahuja_2005}
\BIBentryALTinterwordspacing
A.~T. Ahuja and M.~Ying, ``Sonographic evaluation of cervical lymph nodes,'' \emph{American Journal of Roentgenology}, vol. 184, no.~5, pp. 1691--1699, 2005. [Online]. Available: \url{https://www.ajronline.org/doi/abs/10.2214/ajr.184.5.01841691}
\BIBentrySTDinterwordspacing

\bibitem{cnn_lecun_1989}
Y.~LeCun, B.~Boser, J.~S. Denker, D.~Henderson, R.~E. Howard, W.~Hubbard, and L.~D. Jackel, ``Backpropagation applied to handwritten zip code recognition,'' \emph{Neural computation}, vol.~1, no.~4, pp. 541--551, 1989.

\bibitem{resnet_kaiming_2016}
K.~He, X.~Zhang, S.~Ren, and J.~Sun, ``Deep residual learning for image recognition,'' in \emph{Proceedings of the IEEE conference on computer vision and pattern recognition}, 2016, pp. 770--778.

\bibitem{unet_ronneberger_2015}
O.~Ronneberger, P.~Fischer, and T.~Brox, ``U-net: Convolutional networks for biomedical image segmentation,'' in \emph{Medical image computing and computer-assisted intervention--MICCAI 2015: 18th international conference, Munich, Germany, October 5-9, 2015, proceedings, part III 18}.\hskip 1em plus 0.5em minus 0.4em\relax Springer, 2015, pp. 234--241.

\bibitem{vit_dosovitskiy_2021}
A.~Dosovitskiy, L.~Beyer, A.~Kolesnikov, D.~Weissenborn, X.~Zhai, T.~Unterthiner, M.~Dehghani, M.~Minderer, G.~Heigold, S.~Gelly, J.~Uszkoreit, and N.~Houlsby, ``An image is worth 16x16 words: Transformers for image recognition at scale,'' \emph{ICLR}, 2021.

\bibitem{lesion_cls_adar_2018}
M.~Frid-Adar, I.~Diamant, E.~Klang, M.~Amitai, J.~Goldberger, and H.~Greenspan, ``Gan-based synthetic medical image augmentation for increased cnn performance in liver lesion classification,'' \emph{Neurocomputing}, vol. 321, pp. 321--331, 2018.

\bibitem{vessel_seg_jin_2019}
Q.~Jin, Z.~Meng, T.~D. Pham, Q.~Chen, L.~Wei, and R.~Su, ``Dunet: A deformable network for retinal vessel segmentation,'' \emph{Knowledge-Based Systems}, vol. 178, pp. 149--162, 2019.

\bibitem{practical_ansari_2022}
\BIBentryALTinterwordspacing
M.~Y. Ansari, A.~Abdalla, M.~Y. Ansari, M.~I. Ansari, B.~Malluhi, S.~Mohanty, S.~Mishra, S.~S. Singh, J.~Abinahed, A.~Al-Ansari, S.~Balakrishnan, and S.~P. Dakua, ``Practical utility of liver segmentation methods in clinical surgeries and interventions,'' \emph{BMC Medical Imaging}, vol.~22, no.~1, p.~97, 2022. [Online]. Available: \url{https://doi.org/10.1186/s12880-022-00825-2}
\BIBentrySTDinterwordspacing

\bibitem{neural_ansari_2023}
M.~Y. Ansari, I.~A.~C. Mangalote, D.~Masri, and S.~P. Dakua, ``Neural network-based fast liver ultrasound image segmentation,'' in \emph{2023 International Joint Conference on Neural Networks (IJCNN)}, 2023, Conference Proceedings, pp. 1--8.

\bibitem{towards_ansari_2023}
M.~Y. Ansari, S.~Mohanty, S.~J. Mathew, S.~Mishra, S.~S. Singh, J.~Abinahed, A.~Al-Ansari, and S.~P. Dakua, ``Towards developing a lightweight neural network for liver ct segmentation,'' in \emph{Medical Imaging and Computer-Aided Diagnosis}, R.~Su, Y.~Zhang, H.~Liu, and A.~F~Frangi, Eds.\hskip 1em plus 0.5em minus 0.4em\relax Springer Nature Singapore, 2023, Conference Proceedings, pp. 27--35.

\bibitem{retinaunet_lin_2017}
P.~F. Jaeger, S.~A.~A. Kohl, S.~Bickelhaupt, F.~Isensee, T.~A. Kuder, H.-P. Schlemmer, and K.~H. Maier-Hein, ``Retina u-net: Embarrassingly simple exploitation of segmentation supervision for medical object detection,'' in \emph{Proceedings of the Machine Learning for Health NeurIPS Workshop}, ser. Proceedings of Machine Learning Research, vol. 116.\hskip 1em plus 0.5em minus 0.4em\relax PMLR, 13 Dec 2020, pp. 171--183.

\bibitem{competitive_shan_2019}
H.~Shan, A.~Padole, F.~Homayounieh, U.~Kruger, R.~D. Khera, C.~Nitiwarangkul, M.~K. Kalra, and G.~Wang, ``Competitive performance of a modularized deep neural network compared to commercial algorithms for low-dose ct image reconstruction,'' \emph{Nature Machine Intelligence}, vol.~1, no.~6, pp. 269--276, 2019.

\bibitem{unveiling_ansari_2023}
M.~Y. Ansari, M.~Qaraqe, R.~Righetti, E.~Serpedin, and K.~Qaraqe, ``Unveiling the future of breast cancer assessment: a critical review on generative adversarial networks in elastography ultrasound,'' \emph{Frontiers in Oncology}, vol.~13, 2023.

\bibitem{advancements_ansari_2024}
M.~Y. Ansari, I.~A.~C. Mangalote, P.~K. Meher, O.~Aboumarzouk, A.~Al-Ansari, O.~Halabi, and S.~P. Dakua, ``Advancements in deep learning for b-mode ultrasound segmentation: A comprehensive review,'' \emph{IEEE Transactions on Emerging Topics in Computational Intelligence}, vol.~8, no.~3, pp. 2126--2149, 2024.

\bibitem{deep_rayed_2024}
M.~E. Rayed, S.~M.~S. Islam, S.~I. Niha, J.~R. Jim, M.~M. Kabir, and M.~F. Mridha, ``Deep learning for medical image segmentation: State-of-the-art advancements and challenges,'' \emph{Informatics in Medicine Unlocked}, vol.~47, p. 101504, 2024.

\bibitem{review_aljabri_2022}
M.~Aljabri and M.~AlGhamdi, ``A review on the use of deep learning for medical images segmentation,'' \emph{Neurocomputing}, vol. 506, pp. 311--335, 2022.

\bibitem{prisma_2020}
\BIBentryALTinterwordspacing
M.~J. Page, J.~E. McKenzie, P.~M. Bossuyt, I.~Boutron, T.~C. Hoffmann, C.~D. Mulrow, L.~Shamseer, J.~M. Tetzlaff, E.~A. Akl, S.~E. Brennan, R.~Chou, J.~Glanville, J.~M. Grimshaw, A.~Hróbjartsson, M.~M. Lalu, T.~Li, E.~W. Loder, E.~Mayo-Wilson, S.~McDonald, L.~A. McGuinness, L.~A. Stewart, J.~Thomas, A.~C. Tricco, V.~A. Welch, P.~Whiting, and D.~Moher, ``The prisma 2020 statement: An updated guideline for reporting systematic reviews,'' \emph{International Journal of Surgery}, vol.~88, p. 105906, 2021. [Online]. Available: \url{https://www.sciencedirect.com/science/article/pii/S1743919121000406}
\BIBentrySTDinterwordspacing

\bibitem{lenet_lecun_1998}
Y.~LeCun, L.~Bottou, Y.~Bengio, and P.~Haffner, ``Gradient-based learning applied to document recognition,'' \emph{Proceedings of the IEEE}, vol.~86, no.~11, pp. 2278--2324, 1998.

\bibitem{alexnet_krizhevsky_2017}
A.~Krizhevsky, I.~Sutskever, and G.~E. Hinton, ``Imagenet classification with deep convolutional neural networks,'' \emph{Commun. ACM}, vol.~60, no.~6, pp. 84--90, 2017.

\bibitem{fcn_long_2015}
J.~Long, E.~Shelhamer, and T.~Darrell, ``Fully convolutional networks for semantic segmentation,'' in \emph{Proceedings of the IEEE conference on computer vision and pattern recognition}, 2015, pp. 3431--3440.

\bibitem{holistically_nogues_2016}
I.~Nogues, L.~Lu, X.~Wang, H.~Roth, G.~Bertasius, N.~Lay, J.~Shi, Y.~Tsehay, and R.~M. Summers, ``Automatic lymph node cluster segmentation using holistically-nested neural networks and structured optimization in ct images,'' in \emph{Lecture Notes in Computer Science (including subseries Lecture Notes in Artificial Intelligence and Lecture Notes in Bioinformatics)}, G.~Unal, S.~Ourselin, L.~Joskowicz, M.~R. Sabuncu, and W.~Wells, Eds., vol. 9901 LNCS.\hskip 1em plus 0.5em minus 0.4em\relax Springer Verlag, 2016, pp. 388--397.

\bibitem{hnn_xie_2015}
S.~Xie and Z.~Tu, ``Holistically-nested edge detection,'' in \emph{Proceedings of the IEEE international conference on computer vision}, 2015, pp. 1395--1403.

\bibitem{coarse_zhang_2016}
Y.~Zhang, M.~T.~C. Ying, L.~Yang, A.~T. Ahuja, and D.~Z. Chen, ``Coarse-to-fine stacked fully convolutional nets for lymph node segmentation in ultrasound images,'' in \emph{2016 IEEE International Conference on Bioinformatics and Biomedicine (BIBM)}, 2016, pp. 443--448.

\bibitem{decompose_zhang_2019}
Y.~Zhang, M.~T.~C. Ying, and D.~Z. Chen, ``Decompose-and-integrate learning for multi-class segmentation in medical images,'' in \emph{Lecture Notes in Computer Science (including subseries Lecture Notes in Artificial Intelligence and Lecture Notes in Bioinformatics)}, D.~Shen, P.~T. Yap, T.~Liu, T.~M. Peters, A.~Khan, L.~H. Staib, C.~Essert, and S.~Zhou, Eds., vol. 11765 LNCS.\hskip 1em plus 0.5em minus 0.4em\relax Springer Science and Business Media Deutschland GmbH, 2019, pp. 641--650.

\bibitem{densevoxnet_yu_2017}
L.~Yu, J.-Z. Cheng, Q.~Dou, X.~Yang, H.~Chen, J.~Qin, and P.-A. Heng, ``Automatic 3d cardiovascular mr segmentation with densely-connected volumetric convnets,'' in \emph{Medical Image Computing and Computer-Assisted Intervention- MICCAI 2017: 20th International Conference, Quebec City, QC, Canada, September 11-13, 2017, Proceedings, Part II 20}.\hskip 1em plus 0.5em minus 0.4em\relax Springer, 2017, pp. 287--295.

\bibitem{cumednet_chen_2016}
H.~Chen, X.~Qi, J.-Z. Cheng, and P.-A. Heng, ``Deep contextual networks for neuronal structure segmentation,'' in \emph{Proceedings of the Thirtieth AAAI Conference on Artificial Intelligence}, 2016, pp. 1167--1173.

\bibitem{npcnet_li_2022}
Y.~Li, T.~Dan, H.~Li, J.~Chen, H.~Peng, L.~Liu, and H.~Cai, ``Npcnet: Jointly segment primary nasopharyngeal carcinoma tumors and metastatic lymph nodes in mr images,'' \emph{IEEE Transactions on Medical Imaging}, vol.~41, no.~7, pp. 1639--1650, 2022.

\bibitem{imagenet_deng_2009}
J.~Deng, W.~Dong, R.~Socher, L.-J. Li, K.~Li, and L.~Fei-Fei, ``Imagenet: A large-scale hierarchical image database,'' in \emph{2009 IEEE conference on computer vision and pattern recognition}.\hskip 1em plus 0.5em minus 0.4em\relax Ieee, 2009, pp. 248--255.

\bibitem{ddnn_men_2017}
K.~Men, X.~Chen, Y.~Zhang, T.~Zhang, J.~Dai, J.~Yi, and Y.~Li, ``Deep deconvolutional neural network for target segmentation of nasopharyngeal cancer in planning computed tomography images,'' \emph{Frontiers in oncology}, vol.~7, p. 315, 2017.

\bibitem{vgg_simonyan_2014}
K.~Simonyan and A.~Zisserman, ``Very deep convolutional networks for large-scale image recognition,'' \emph{arXiv preprint arXiv:1409.1556}, 2014.

\bibitem{tumor_li_2019}
S.~Li, J.~Xiao, L.~He, X.~Peng, and X.~Yuan, ``The tumor target segmentation of nasopharyngeal cancer in ct images based on deep learning methods,'' \emph{Technology in cancer research \& treatment}, vol.~18, p. 1533033819884561, 2019.

\bibitem{seg_ariji_2022}
Y.~Ariji, Y.~Kise, M.~Fukuda, C.~Kuwada, and E.~Ariji, ``Segmentation of metastatic cervical lymph nodes from ct images of oral cancers using deep-learning technology,'' \emph{Dentomaxillofac Radiol}, vol.~51, no.~4, p. 20210515, 2022.

\bibitem{mediastinal_nayan_2022}
A.~A. Nayan, B.~Kijsirikul, and Y.~Iwahori, ``Mediastinal lymph node detection and segmentation using deep learning,'' \emph{IEEE Access}, vol.~10, pp. 89\,289--89\,307, 2022.

\bibitem{unet++_zhou_2018}
Z.~Zhou, M.~M. Rahman~Siddiquee, N.~Tajbakhsh, and J.~Liang, ``Unet++: A nested u-net architecture for medical image segmentation,'' in \emph{Deep Learning in Medical Image Analysis and Multimodal Learning for Clinical Decision Support: 4th International Workshop, DLMIA 2018, and 8th International Workshop, ML-CDS 2018, Held in Conjunction with MICCAI 2018, Granada, Spain, September 20, 2018, Proceedings 4}.\hskip 1em plus 0.5em minus 0.4em\relax Springer, 2018, pp. 3--11.

\bibitem{disegnet_xu_2021}
G.~Xu, H.~Cao, J.~K. Udupa, Y.~Tong, and D.~A. Torigian, ``Disegnet: A deep dilated convolutional encoder-decoder architecture for lymph node segmentation on pet/ct images,'' \emph{Comput Med Imaging Graph}, vol.~88, p. 101851, 2021.

\bibitem{segnet_badrinarayanan_2017}
V.~Badrinarayanan, A.~Kendall, and R.~Cipolla, ``Segnet: A deep convolutional encoder-decoder architecture for image segmentation,'' \emph{IEEE transactions on pattern analysis and machine intelligence}, vol.~39, no.~12, pp. 2481--2495, 2017.

\bibitem{seg_ahamed_2023}
S.~Ahamed, L.~Polson, and A.~Rahmim, ``A u-net convolutional neural network with multiclass dice loss for automated segmentation of tumors and lymph nodes from head and neck cancer pet/ct images,'' in \emph{Head and Neck Tumor Segmentation and Outcome Prediction}.\hskip 1em plus 0.5em minus 0.4em\relax Springer Nature Switzerland, 2023, pp. 94--106.

\bibitem{multi_modal_fu_2020}
X.~Fu, T.~Gao, Y.~Liu, M.~Zhang, C.~Guo, J.~Wu, and Z.~Wang, ``Multi-modal feature attention for cervical lymph node segmentation in ultrasound and doppler images,'' in \emph{Communications in Computer and Information Science}, H.~Yang, K.~Pasupa, A.~C. Leung, J.~T. Kwok, J.~H. Chan, and I.~King, Eds., vol. 1332.\hskip 1em plus 0.5em minus 0.4em\relax Springer Science and Business Media Deutschland GmbH, 2020, pp. 479--487.

\bibitem{transformer_vaswani_2017}
A.~Vaswani, N.~Shazeer, N.~Parmar, J.~Uszkoreit, L.~Jones, A.~N. Gomez, {\L}.~Kaiser, and I.~Polosukhin, ``Attention is all you need,'' \emph{Advances in neural information processing systems}, vol.~30, 2017.

\bibitem{ssd_liu_2016}
W.~Liu, D.~Anguelov, D.~Erhan, C.~Szegedy, S.~Reed, C.-Y. Fu, and A.~C. Berg, ``Ssd: Single shot multibox detector,'' in \emph{Computer Vision--ECCV 2016: 14th European Conference, Amsterdam, The Netherlands, October 11--14, 2016, Proceedings, Part I 14}.\hskip 1em plus 0.5em minus 0.4em\relax Springer, 2016, pp. 21--37.

\bibitem{manet_zhang_2020}
L.~Zhang, J.~Zhang, Z.~Li, and Y.~Song, ``A multiple-channel and atrous convolution network for ultrasound image segmentation,'' \emph{Med Phys}, vol.~47, no.~12, pp. 6270--6285, 2020.

\bibitem{munet_zhang_2020}
W.~Zhang, H.~Cheng, and J.~Gan, ``Munet: A multi-scale u-net framework for medical image segmentation,'' in \emph{2020 International Joint Conference on Neural Networks (IJCNN)}.\hskip 1em plus 0.5em minus 0.4em\relax IEEE, 2020, pp. 1--7.

\bibitem{yolov3_redmon_2018}
J.~Redmon and A.~Farhadi, ``Yolov3: An incremental improvement,'' \emph{arXiv preprint arXiv:1804.02767}, 2018.

\bibitem{diffusion_chen_2021}
H.~Chen, Y.~Wang, J.~Shi, J.~Xiong, J.~Jiang, W.~Chang, M.~Chen, and Q.~Zhang, ``Segmentation of lymph nodes in ultrasound images using u-net convolutional neural networks and gabor-based anisotropic diffusion,'' \emph{Journal of Medical and Biological Engineering}, vol.~41, no.~6, pp. 942--952, 2021.

\bibitem{difficulty_xu_2022}
Q.~Xu, X.~Xi, X.~Meng, Z.~Qin, X.~Nie, Y.~Wu, D.~Zhou, Y.~Qu, C.~Li, and Y.~Yin, ``Difficulty-aware bi-network with spatial attention constrained graph for axillary lymph node segmentation,'' \emph{Science China Information Sciences}, vol.~65, no.~9, 2022.

\bibitem{deeplabv3+_chen_2018}
L.-C. Chen, Y.~Zhu, G.~Papandreou, F.~Schroff, and H.~Adam, ``Encoder-decoder with atrous separable convolution for semantic image segmentation,'' in \emph{Proceedings of the European conference on computer vision (ECCV)}, 2018, pp. 801--818.

\bibitem{frrn_pohlen_2017}
T.~Pohlen, A.~Hermans, M.~Mathias, and B.~Leibe, ``Full-resolution residual networks for semantic segmentation in street scenes,'' in \emph{Proceedings of the IEEE conference on computer vision and pattern recognition}, 2017, pp. 4151--4160.

\bibitem{efficient_wen_2024}
F.~Wen, J.~Zhou, Z.~B. Chen, M.~Dou, Y.~Yao, X.~Wang, F.~Xu, and Y.~L. Shen, ``Efficient application of deep learning-based elective lymph node regions delineation for pelvic malignancies,'' \emph{MEDICAL PHYSICS}, vol.~51, no.~10, pp. 7057--7066, 2024.

\bibitem{deep_zhao_2024}
H.~N. Zhao, H.~Yin, J.~Y. Liu, L.~L. Song, Y.~L. Peng, and B.~Y. Ma, ``Deep learning-assisted ultrasonic diagnosis of cervical lymph node metastasis of thyroid cancer: a retrospective study of 3059 patients,'' \emph{Frontiers in Oncology}, vol.~14, 2024.

\bibitem{automated_hasan_2024}
M.~M. Al~Hasan, S.~Ghazimoghadam, P.~Tunlayadechanont, M.~T. Mostafiz, M.~Gupta, A.~Roy, K.~Peters, B.~Hochhegger, A.~Mancuso, N.~Asadizanjani, and R.~Forghani, ``Automated segmentation of lymph nodes on neck ct scans using deep learning,'' \emph{J Imaging Inform Med}, vol.~37, no.~6, pp. 2955--2966, 2024.

\bibitem{swint_liu_2021}
Z.~Liu, Y.~Lin, Y.~Cao, H.~Hu, Y.~Wei, Z.~Zhang, S.~Lin, and B.~Guo, ``Swin transformer: Hierarchical vision transformer using shifted windows,'' in \emph{Proceedings of the IEEE/CVF international conference on computer vision}, 2021, pp. 10\,012--10\,022.

\bibitem{denet_shi_2022}
J.~Shi, Z.~Wang, H.~Kan, M.~Zhao, X.~Xue, B.~Yan, H.~An, J.~Shen, J.~Bartlett, W.~Lu, and J.~Duan, ``Automatic segmentation of target structures for total marrow and lymphoid irradiation in bone marrow transplantation,'' in \emph{2022 44th Annual International Conference of the IEEE Engineering in Medicine \& Biology Society (EMBC)}, 2022, pp. 5025--5029.

\bibitem{utnet_gao_2021}
Y.~Gao, M.~Zhou, and D.~N. Metaxas, ``Utnet: a hybrid transformer architecture for medical image segmentation,'' in \emph{Medical Image Computing and Computer Assisted Intervention--MICCAI 2021: 24th International Conference, Strasbourg, France, September 27--October 1, 2021, Proceedings, Part III 24}.\hskip 1em plus 0.5em minus 0.4em\relax Springer, 2021, pp. 61--71.

\bibitem{medvit_manzari_2023}
O.~N. Manzari, H.~Ahmadabadi, H.~Kashiani, S.~B. Shokouhi, and A.~Ayatollahi, ``Medvit: a robust vision transformer for generalized medical image classification,'' \emph{Computers in biology and medicine}, vol. 157, p. 106791, 2023.

\bibitem{efficientvit_cai_2023}
H.~Cai, J.~Li, M.~Hu, C.~Gan, and S.~Han, ``Efficientvit: Lightweight multi-scale attention for high-resolution dense prediction,'' in \emph{Proceedings of the IEEE/CVF International Conference on Computer Vision}, 2023, pp. 17\,302--17\,313.

\bibitem{mobilevit_mehta_2022}
S.~Mehta and M.~Rastegari, ``Mobilevit: Light-weight, general-purpose, and mobile-friendly vision transformer,'' in \emph{International Conference on Learning Representations}, 2022.

\bibitem{semantic_bouget_2019}
D.~Bouget, A.~J{\o}rgensen, G.~Kiss, H.~O. Leira, and T.~Lang{\o}, ``Semantic segmentation and detection of mediastinal lymph nodes and anatomical structures in ct data for lung cancer staging,'' \emph{International journal of computer assisted radiology and surgery}, vol.~14, pp. 977--986, 2019.

\bibitem{mri_zhao_2020}
X.~Zhao, P.~Xie, M.~Wang, W.~Li, P.~J. Pickhardt, W.~Xia, F.~Xiong, R.~Zhang, Y.~Xie, J.~Jian, H.~Bai, C.~Ni, J.~Gu, T.~Yu, Y.~Tang, X.~Gao, and X.~Meng, ``Deep learning-based fully automated detection and segmentation of lymph nodes on multiparametric-mri for rectal cancer: A multicentre study,'' \emph{EBioMedicine}, vol.~56, p. 102780, 2020.

\bibitem{maskrcnn_he_2017}
K.~He, G.~Gkioxari, P.~Doll{\'a}r, and R.~Girshick, ``Mask r-cnn,'' in \emph{Proceedings of the IEEE international conference on computer vision}, 2017, pp. 2961--2969.

\bibitem{yolov8_reis_2023}
D.~Reis, J.~Kupec, J.~Hong, and A.~Daoudi, ``Real-time flying object detection with yolov8,'' \emph{arXiv preprint arXiv:2305.09972}, 2023.

\bibitem{fasterrcnn_ren_2016}
S.~Ren, K.~He, R.~Girshick, and J.~Sun, ``Faster r-cnn: Towards real-time object detection with region proposal networks,'' \emph{IEEE transactions on pattern analysis and machine intelligence}, vol.~39, no.~6, pp. 1137--1149, 2016.

\bibitem{detr_carion_2020}
N.~Carion, F.~Massa, G.~Synnaeve, N.~Usunier, A.~Kirillov, and S.~Zagoruyko, ``End-to-end object detection with transformers,'' in \emph{European conference on computer vision}.\hskip 1em plus 0.5em minus 0.4em\relax Springer, 2020, pp. 213--229.

\bibitem{focal_xu_2020}
G.~Xu, H.~Cao, Y.~Dong, C.~Yue, K.~Li, and Y.~Tong, ``Focal loss function based deeplabv3+ for pathological lymph node segmentation on pet/ct,'' in \emph{ACM International Conference Proceeding Series}.\hskip 1em plus 0.5em minus 0.4em\relax Association for Computing Machinery, 2020, pp. 24--28.

\bibitem{atte_loss_xu_2020}
G.~Xu, H.~Cao, and G.~Jiang, ``Boundary-attention loss function in neural network for pathological lymph nodes segmentation based on pet/ct images,'' in \emph{ACM International Conference Proceeding Series}.\hskip 1em plus 0.5em minus 0.4em\relax Association for Computing Machinery, 2020, pp. 90--94.

\bibitem{tcia_roth_2018}
H.~Roth, L.~Le, S.~Ari, K.~Cherry, J.~Hoffman, S.~Wang, and R.~Summers, ``A new 2.5 d representation for lymph node detection in ct,'' \emph{The Cancer Imaging Archive}, 2018.

\bibitem{deconv_noh_2015}
H.~Noh, S.~Hong, and B.~Han, ``Learning deconvolution network for semantic segmentation,'' in \emph{Proceedings of the IEEE international conference on computer vision}, 2015, pp. 1520--1528.

\bibitem{tcia_li_2020}
\BIBentryALTinterwordspacing
P.~Li, S.~Wang, T.~Li, J.~Lu, Y.~HuangFu, and D.~Wang, ``A large-scale ct and pet/ct dataset for lung cancer diagnosis (lung-pet-ct-dx),'' \emph{The Cancer Imaging Archive}, 2020. [Online]. Available: \url{https://doi.org/10.7937/TCIA.2020.NNC2-0461}
\BIBentrySTDinterwordspacing

\bibitem{5-patients_2022}
C.~Laboratory, ``Atlas of mediastinal lymph stations,'' \url{https://www.creatis.insa-lyon.fr/lymph-stations-atlas/data/14a2770a2e.php}, 2022.

\bibitem{lung_armato_2015}
S.~Armato, ``Public lung image databases,'' in \emph{Computer-Aided Detection and Diagnosis in Medical Imaging}.\hskip 1em plus 0.5em minus 0.4em\relax CRC Press, 2015, pp. 218--229.

\bibitem{hecktor_2022}
T.~M. Society, ``Hecktor 2022 - grand challenge,'' \url{https://hecktor.grand-challenge.org/Data/}, 2022.

\bibitem{dilated_yu_2015}
F.~Yu and V.~Koltun, ``Multi-scale context aggregation by dilated convolutions,'' \emph{arXiv preprint arXiv:1511.07122}, 2015.

\bibitem{inceptionv3_szegedy_2016}
C.~Szegedy, V.~Vanhoucke, S.~Ioffe, J.~Shlens, and Z.~Wojna, ``Rethinking the inception architecture for computer vision,'' in \emph{Proceedings of the IEEE conference on computer vision and pattern recognition}, 2016, pp. 2818--2826.

\bibitem{aspp_chen_2017}
L.-C. Chen, G.~Papandreou, F.~Schroff, and H.~Adam, ``Rethinking atrous convolution for semantic image segmentation,'' \emph{arXiv preprint arXiv:1706.05587}, 2017.

\bibitem{focusnet_gao_2019}
Y.~Gao, R.~Huang, M.~Chen, Z.~Wang, J.~Deng, Y.~Chen, Y.~Yang, J.~Zhang, C.~Tao, and H.~Li, ``Focusnet: Imbalanced large and small organ segmentation with an end-to-end deep neural network for head and neck ct images,'' in \emph{Medical Image Computing and Computer Assisted Intervention - MICCAI 2019}, D.~Shen, T.~Liu, T.~M. Peters, L.~H. Staib, C.~Essert, S.~Zhou, P.-T. Yap, and A.~Khan, Eds.\hskip 1em plus 0.5em minus 0.4em\relax Springer International Publishing, 2019, Conference Proceedings, pp. 829--838.

\bibitem{airway_reynisson_2015}
P.~J. Reynisson, M.~Scali, E.~Smistad, E.~F. Hofstad, H.~O. Leira, F.~Lindseth, T.~A. Nagelhus~Hernes, T.~Amundsen, H.~Sorger, and T.~Lang{\o}, ``Airway segmentation and centerline extraction from thoracic ct--comparison of a new method to state of the art commercialized methods,'' \emph{PloS one}, vol.~10, no.~12, p. e0144282, 2015.

\bibitem{continuous_wang_2024}
L.~Wang, X.~Zhang, H.~Su, and J.~Zhu, ``A comprehensive survey of continual learning: Theory, method and application,'' \emph{IEEE Transactions on Pattern Analysis and Machine Intelligence}, vol.~46, no.~8, pp. 5362--5383, 2024.

\bibitem{coco_lin_2014}
T.-Y. Lin, M.~Maire, S.~Belongie, J.~Hays, P.~Perona, D.~Ramanan, P.~Doll{\'a}r, and C.~L. Zitnick, ``Microsoft coco: Common objects in context,'' in \emph{Computer Vision--ECCV 2014: 13th European Conference, Zurich, Switzerland, September 6-12, 2014, Proceedings, Part V 13}.\hskip 1em plus 0.5em minus 0.4em\relax Springer, 2014, pp. 740--755.

\bibitem{cityscapes_cordts_2016}
M.~Cordts, M.~Omran, S.~Ramos, T.~Rehfeld, M.~Enzweiler, R.~Benenson, U.~Franke, S.~Roth, and B.~Schiele, ``The cityscapes dataset for semantic urban scene understanding,'' in \emph{Proceedings of the IEEE conference on computer vision and pattern recognition}, 2016, pp. 3213--3223.

\bibitem{sam_dataset_ye_2023}
J.~Ye, J.~Cheng, J.~Chen, Z.~Deng, T.~Li, H.~Wang, Y.~Su, Z.~Huang, J.~Chen, L.~Jiang \emph{et~al.}, ``Sa-med2d-20m dataset: Segment anything in 2d medical imaging with 20 million masks,'' \emph{arXiv preprint arXiv:2311.11969}, 2023.

\bibitem{bcp_bai_2023}
Y.~Bai, D.~Chen, Q.~Li, W.~Shen, and Y.~Wang, ``Bidirectional {Copy}-{Paste} for {Semi}-{Supervised} {Medical} {Image} {Segmentation}.'' in \emph{Computer {Vision} and {Pattern} {Recognition} ({CVPR})}, 2023, pp. 11\,514--11\,524.

\bibitem{ccam_chen_2022}
Z.~Chen, Z.~Tian, J.~Zhu, C.~Li, and S.~Du, ``C-cam: Causal cam for weakly supervised semantic segmentation on medical image,'' in \emph{Proceedings of the IEEE/CVF Conference on Computer Vision and Pattern Recognition (CVPR)}, June 2022, pp. 11\,676--11\,685.

\bibitem{byol_grill_2020}
J.-B. Grill, F.~Strub, F.~Altch{\'e}, C.~Tallec, P.~Richemond, E.~Buchatskaya, C.~Doersch, B.~Avila~Pires, Z.~Guo, M.~Gheshlaghi~Azar \emph{et~al.}, ``Bootstrap your own latent-a new approach to self-supervised learning,'' \emph{Advances in neural information processing systems}, vol.~33, pp. 21\,271--21\,284, 2020.

\bibitem{unsupervised_chen_2020}
C.~Chen, Q.~Dou, H.~Chen, J.~Qin, and P.~A. Heng, ``Unsupervised bidirectional cross-modality adaptation via deeply synergistic image and feature alignment for medical image segmentation,'' \emph{IEEE Transactions on Medical Imaging}, vol.~39, no.~7, pp. 2494--2505, 2020.

\bibitem{tiny_vit_wu_2022}
K.~Wu, J.~Zhang, H.~Peng, M.~Liu, B.~Xiao, J.~Fu, and L.~Yuan, ``Tinyvit: Fast pretraining distillation for small vision transformers,'' in \emph{European conference on computer vision (ECCV)}, 2022.

\bibitem{weighted_mahdi_2024}
M.~A. Mahdi, S.~Ahamad, S.~A. Saad, A.~Dafhalla, R.~Qureshi, and A.~Alqushaibi, ``Weighted fusion transformer for dual pet/ct head and neck tumor segmentation,'' \emph{IEEE Access}, vol.~12, pp. 110\,905--110\,919, 2024.

\bibitem{gpt_brown_2020}
T.~Brown, B.~Mann, N.~Ryder, M.~Subbiah, J.~D. Kaplan, P.~Dhariwal, A.~Neelakantan, P.~Shyam, G.~Sastry, A.~Askell \emph{et~al.}, ``Language models are few-shot learners,'' \emph{Advances in neural information processing systems}, vol.~33, pp. 1877--1901, 2020.

\bibitem{sam_kirillov_2023}
A.~Kirillov, E.~Mintun, N.~Ravi, H.~Mao, C.~Rolland, L.~Gustafson, T.~Xiao, S.~Whitehead, A.~C. Berg, W.-Y. Lo \emph{et~al.}, ``Segment anything,'' in \emph{Proceedings of the IEEE/CVF International Conference on Computer Vision}, 2023, pp. 4015--4026.

\bibitem{sam_med2d_cheng_2023}
J.~Cheng, J.~Ye, Z.~Deng, J.~Chen, T.~Li, H.~Wang, Y.~Su, Z.~Huang, J.~Chen, L.~Jiang \emph{et~al.}, ``Sam-med2d,'' \emph{arXiv preprint arXiv:2308.16184}, 2023.

\bibitem{collaborative_wu_2024}
S.-H. Wu, W.-J. Tong, M.-D. Li, H.-T. Hu, X.-Z. Lu, Z.-R. Huang, X.-X. Lin, R.-F. Lu, M.-D. Lu, L.-D. Chen \emph{et~al.}, ``Collaborative enhancement of consistency and accuracy in us diagnosis of thyroid nodules using large language models,'' \emph{Radiology}, vol. 310, no.~3, p. e232255, 2024.

\end{thebibliography}

\end{document}